\documentclass[aps, 10pt, prl,notitlepage, twocolumn, superscriptaddress,nofootinbib]{revtex4-1}

\usepackage{amsmath}
\usepackage{amssymb}
\usepackage{amsfonts}
\usepackage[utf8]{inputenc}
\usepackage[T1]{fontenc}
\usepackage{mathrsfs}
\usepackage{anyfontsize}

\usepackage[usenames,dvipsnames]{xcolor}

\usepackage{newtxtext}
\usepackage{newtxmath}
\usepackage{tensor}
\usepackage{bm}

\usepackage{graphicx}
\usepackage{epsfig}
\usepackage{epstopdf}

\usepackage{natbib}

\usepackage[normalem]{ulem}

\allowdisplaybreaks[1]
\begin{document}

\title {New algorithms to obtain analytical solutions of Einstein's equations in isotropic coordinates}

\author{C. Las Heras}
\email{camilo.lasheras@ua.cl}
\affiliation{Departamento de F\'isica, Universidad de Antofagasta,  Aptdo 02800, Chile.}

\author{P. Le\'on}
\email{pablo.leon@ua.cl}
\affiliation{Departamento de F\'isica, Universidad de Antofagasta, Aptdo 02800, Chile.}

%%%%%%%%%%%%%%%%%%%%%%%%%%%%%%%%%%%%  DATE  %%%%%%%%%%%%%%%%%%%%%%%%%%%%%%%%%%%%
%\date{\today}

\begin{abstract}
The main objective of this work, is to show two inequivalent methods to obtain new spherical symmetric solutions of Einstein's Equations with anisotropy in the pressures in isotropic coordinates. This was done inspired by the MGD method, which is known to be valid for line elements in Schwarzschild coordinates. As example, we obtained four analytical solutions using Gold III as seed solution. Two solutions, out of four, (one for each algorithm), satisfy the physical acceptability conditions.
\end{abstract}



\maketitle

\section{Introduction}
\label{intro}
An interesting problem for both, mathematicians and physicists, is to find new analytical solutions of Einstein's equations. This is not an easy task because, in general, it corresponds to a system of coupled differential equations. Moreover, in order to be a realistic solution, it must satisfy certain conditions for physical acceptability. For these reasons, most of the known solutions to Einstein's equations in the literature, are related to specific restrictions about the geometry of the space-time or particular distributions of matter (like the isotropic perfect fluid) \cite{Stephani,lake2,Delgaty}.

Recently, a method was discovered in the context of Randall-Sundrum Brane World \cite{Randall1,Randall2} named Minimal Geometric Deformation (MGD) \cite{Ovalle1,Ovalle2}. It has many different applications\cite{Ovalle4,Ovalle5,Ovalle6,Ovalle7,Ovalle8,Ovalle9,Ovalle10,Ovalle11,Casadio,Darocha1,Darocha2,Darocha3,Contreras,Contreras2,Contreras3,Contreras4,Contreras5,Contreras6} and it allows us to find new analytical solutions to Einstein's equations by decoupling gravitational sources in GR \cite{Ovalle}. This solution could describe the inner and outer regions of compact objects, black holes and the coupling of the Einstein's equation with a scalar field, among many other possibilities. It is important to mention that these solutions are in general anisotropic in the pressures. This is really important when dealing with an internal solution. There are many theoretical reasons to consider local anisotropy of the pressures \cite{Herrera2,Herrera3,Herrera4,Herrera5} in order to describe more realistic gravitational objects. In \cite{lhl} we found 2 new families of anisotropic solutions, and we obtained an interval of values where the free parameters shall dwell in order to reproduce a physically acceptable solution. There are a good number of references that shows the efficiency of this method to obtain useful solutions to describe realistic systems that considers anisotropy in the pressures \cite{Gabbanelli,Tello1,Morales1,Morales2,Estrada1,Estrada2}. However, MGD requires spherical symmetry, and considers solutions of Einstein's equations whose line element, is written in Schwarzschild's coordinates (also known by the name of canonical coordinates).

Nevertheless, there are other known exact solutions of Einstein's equations written in other coordinates systems. In some of these cases, the transformation to move to the Schwarzschild coordinates is not always well defined. This becomes evident when we work in isotropic coordinates. It has been shown (see for example \cite{Nariai}) that every line element given in Schwarzschild-like coordinates can be transformed into the isotropic one, but the reverse process is not always possible. This seems to indicate that the isotropic coordinates are more general that canonical coordinates. The main feature of the isotropic line elements is that it treats the 3 spatial dimensions in the same way. Also, this system of coordinates has been adopted by physicist and astronomers, over the Schwarzschild coordinates in order to describe the solar system \cite{Green}. 

In this work we present two inequivalent algorithms to obtain new analytical and anisotropic solutions of Einstein's equations, starting with known solutions in isotropic coordinates. These two algorithms are based in the same ansatz of the MGD method. However, there is no decoupling of Einstein's equation in the first one, while in the second one, it is necessary to assume additional conditions that are not included in the MGD aproach. In order to check the validity of the methods, we have chosen a known seed solution in isotropic coordinates, as Gold III \cite{Gold}. We have founded four different analytical solutions. A few different algorithms that also allow us to obtain new solutions from a line element in isotropic coordinates are explained in \cite{Maharaj,Maharaj2,Maharaj3}. However, our method is simpler and it will lead us, generally, to solutions with anisotropy in the pressures, which is not the case in \cite{Maharaj}.  

This paper is organized as follows: in section 2 we briefly review Einstein's equations in standard coordinates. In section 3 we describe the solution-generating technique known as MGD, which is useful with line elements in Schwarzschild-like coordinates. We briefly discuss Einstein's equations on isotropic coordinates in section 4. The acceptability conditions for the solutions of Einstein's equations are presented in the section 5. In section 6 we proposed two different, MGD inspired, algorithms that allow us to obtain new analytical solutions in isotropic coordinates, with anisotropy in the pressures. In order to verify the usefulness of the methods, we extend the Gold III solution to the anisotropical domain with both methods and using two different conditions. The physical acceptability conditions of the four new and exact anisotropic solutions are also discussed in Section 7. 

\section{Einstein's equations in standard coordinates}
\label{sec:1}

The line element in standard  coordinates, also known as Schwarzschild like coordinates, is given by
\begin{equation}
ds^{2}=e^{\nu (r)}\,dt^{2}-\frac{1}{\mu(r)}\,dr^{2}-r^{2}\left( d\theta^{2}+\sin ^{2}\theta \,d\phi ^{2}\right).
\label{2.1}
\end{equation}
This metric must be a solution of Einstein's equations
\begin{equation}
\label{2.2}
R_{\mu\nu}-\frac{1}{2}\,R\, g_{\mu\nu}=-8\pi\,T^._{\mu\nu},
\end{equation}
which take the following form 
\begin{eqnarray}
\label{2.3}
8\pi T_0^0 &=&\frac{1}{r^2}-\frac{\mu}{r^2}-\frac{\mu'}{r}\ ,\\
\label{2.4}
-8\pi T_1^1 &=&-\frac 1{r^2}+\mu\left( \frac 1{r^2}+\frac{\nu'}r\right)\ ,\\
\label{2.5}
-8\pi T_2^2&=&\frac{\mu}{4}\left(2\nu''+\nu'^2+\frac{2\nu'}{r}\right)+\frac{\mu'}{4}\left(\nu'+\frac{2}{r}\right),
\end{eqnarray}
where the prime indicates derivatives respect to variable $r$.

Finally the conservation equation, derived from the latter system of equations is
\begin{equation}
\label{2.6}
\nabla_\mu T^{\mu\nu}=0,
\end{equation}
whose radial component lead to the equilibrium equation of the matter distribution

\begin{equation}
    8\pi (T_1^1)' +\frac{16\pi}{r}(T_1^1-T_2^2)+4\pi \nu' (T_1^1-T_0^0)=0
\end{equation}

%where the perfect fluid case is recovered for $\alpha= 0$.

%We notice that the system (\ref{2.5}-\ref{2.7}) has seven unknown functions ($\lambda(r)$ , $\nu(r)$ , $p(r)$ , $\rho(r)$ , $\theta_0^0$ , $\theta_1^1$ , $\theta_2^2$) and three independent equations. The variables can be reduced to five if we identify the effective radial pressure $\tilde{p}_r$, effective tangential pressure $\tilde{p}_t$ and effective density $\tilde{\rho}$.

%Furthermore, there will be in general an anisotropy function given by
%\begin{equation}
%\label{2.9}
%\Pi
%\equiv
%\tilde{p}_{t}-\tilde{p}_{r}
%=
%\alpha\left(\theta_1^{\,1}-\theta_2^{\,2}\right)
%\ ,
%\end{equation}
%inside the stellar distribution.

%We must have more information in order to solve the system, this can be achieved by arbitrarly fixing at least two of the variables, or assuming certain relations between them. It can be done via a state equation or via de MGD approach.

\section{The MGD Approach}
\label{sec:2}

	The starting point of this method is to assume that the energy-momentum tensor has the specific form 
\begin{equation}
\label{3.1}
T_{\mu\nu}=T^{\rm (PF)}_{\mu\nu}+\alpha\,\theta_{\mu\nu},
\end{equation}
where $\alpha$ is a coupling constant and $T^{\rm (PF)}$ is the matter-energy content associated to a perfect fluid 
\begin{equation}
\label{3.2}
T^{\rm (PF)}_{\mu \nu }=(\rho +p)\,u_{\mu }\,u_{\nu }-p\,g_{\mu \nu },
\end{equation}
with the fluid 4-velocity given by $u^{\mu }=e^{-\nu /2}\,\delta _{0}^{\mu }$.  

Now we can consider a perfect fluid solution of Einstein's equations ($\alpha=0$) with a line element written in standard coordinates as (\ref{2.1}), and define 
\begin{equation}
\label{3.3}
\mu(r)\equiv 1-\frac{8\pi}{r}\int_0^r x^2\,\rho\, dx
=1-\frac{2\,m(r)}{r},
\end{equation}
which is the standard expression for the mass function in GR.  
The next step is to take account of the anisotropy introduced by the gravitational source $\theta_{\mu\nu}$ in our system. This will be done by assuming  that the contribution of the $\alpha$ parameter, in the perfect fluid solution (\ref{2.1}), is encoded in the deformations $g$ and $f$ of the temporal and radial metric component respectively.
\begin{eqnarray}
\label{3.4}
\nu
&\mapsto & 
\tilde{\nu}
=
\nu+\alpha\,g
\ ,
\\
\label{3.5}
\mu 
&\mapsto &
\tilde{\mu}
=
\mu+\alpha\,f
\ ,
\end{eqnarray}

The Minimal Geometric Deformation will be associated to the specific case where $g=0$. In this case ($g=0$) is easy to check that, using (\ref{3.1}) and (\ref{3.5}), Einstein's equations~(\ref{2.3})-(\ref{2.5}) splits in two systems. The first one coincides with Einstein's equations system for a perfect fluid 
\begin{eqnarray}
\label{3.6} 
8\pi\rho & = & \frac{1}{r^2} -\frac{\mu}{r^2} -\frac{\mu'}{r}\ , \\
\label{3.7}
8\pi p & = & -\frac 1{r^2}+\mu\left( \frac 1{r^2}+\frac{\nu'}r\right)\ , \\
\label{3.8}
8\pi p & = & \frac{\mu}{4}\left(2\nu''+\nu'^2+\frac{2\nu'}{r}\right)+\frac{\mu'}{4}\left(\nu'+\frac{2}{r}\right) \ ,
\end{eqnarray}
with the correspondent conservation equation \begin{eqnarray}
\label{3.9}
p'+\frac{\nu'}{2}\left(\rho+p\right) = 0,
\end{eqnarray}
that turns out to be equation (\ref{2.6}) with the energy momentum tensor associated to a perfect fluid (\ref{3.2})(eq (\ref{3.1}) with  $\alpha=0$).

The second system of equations, which is not an Einstein's equations system due to the lackness of a $\frac{1}{r}$ on the right side of the first two equations, reads
\begin{eqnarray}
\label{3.10}
8\pi\,\theta_0^{\,0}
&\!\!=\!\!&
-\frac{f^{*}}{r^2}
-\frac{f^{*'}}{r}
\ ,
\\
\label{3.11}
8\pi\,\theta_1^{\,1}
&\!\!=\!\!&
-f^{*}\left(\frac{1}{r^2}+\frac{\nu'}{r}\right)
\ ,
\\
\label{3.12}
8\pi\,\theta_2^{\,2}
&\!\!=\!\!&
-\frac{f^{*}}{4}\left(2\nu''+\nu'^2+2\frac{\nu'}{r}\right)
-\frac{f^{*'}}{4}\left(\nu'+\frac{2}{r}\right)
\ ,
\end{eqnarray}
and the conservation equation associated with the source is
\begin{eqnarray}
\label{3.13}
\left(\theta_1^{\,\,1}\right)'
-\frac{\nu'}{2}\left(\theta_0^{\,\,0}-\theta_1^{\,\,1}\right)
-\frac{2}{r}\left(\theta_2^{\,\,2}-\theta_1^{\,\,1}\right)
=
0.
\end{eqnarray}

In order to find the solution of Einstein's equations for an energy-momentum tensor of the form (\ref{3.1}) we have to solve the equations systems (\ref{3.6})-(\ref{3.8})  and (\ref{3.10})-(\ref{3.12}). In the case when we start with a known perfect fluid solution, then is only necessary to solve the second system. Now, in both cases there are more unknown functions that equations so additional information is required in order to solve the systems. This information can be given in the form of equation of state or other expressions that relates the physical variables of the system under study. 

Finally it can be seen that redefining an energy momentum tensor $\tilde{\theta}_{\mu\nu}$, the system of equations (\ref{3.10})-(\ref{3.12}) associated to the source, is equivalent to a Einstein's equations system for an anisotropic fluid. Also, it can be verified that the conservation equations (\ref{3.9}) and (\ref{3.13}), implies that the interaction of the perfect fluid with the source is purely gravitational. 

\section{Einstein's equations in isotropic coordinates}
\label{sec:3}

In this paper we shall study the interior of static and spherical symmetric matter distributions, but using the line element in isotropic coordinates given by
\begin{eqnarray}\label{4.1}
ds^2=e^{\nu(r)} dt^2-\frac{1}{\omega}(dr^2+r^2d\Omega^2),
\end{eqnarray}
in which Einstein's equations take the following form
\begin{eqnarray}
8\pi T_0^0&=&\omega''-\frac{5}{4}\frac{\omega'^2}{\omega}+\frac{2}{r}\omega' , \label{4.2} \\
-8\pi T_1^1&=&\frac{1}{4}\frac{\omega'^2}{\omega}-\frac{1}{2}\omega'\nu'+\frac{\nu'\omega-\omega'}{r} , \label{4.3}  \\
-8\pi T_2^2 &=&\frac{1}{2}\left( \frac{\omega'^2}{\omega}-\omega''\right)+\left( \frac{\nu''}{2}+\frac{\nu'^2}{4}+\frac{\nu'}{2r}\right)\omega \nonumber \\ & - & \frac{\omega'}{2r}.  \label{4.4}
\end{eqnarray}

This coordinates seems to be more general than the Schwarzschild like  ones (\ref{2.1}), due to the fact that there is always possible to transform the line element from the standard form to the isotropic one by 
\begin{eqnarray}\label{4.5}
r=A\exp \left\lbrace  \int \mu^{-1/2}\frac{dr_1}{r_1}\right\rbrace ,
\end{eqnarray}
where $\mu^{-1}$ and $r_1$ are the radial component and the coordinate associated to the line element in standard coordinates (\ref{2.1}). However, is not always possible to perform the reverse process. Therefore, there is a chance to obtain solutions to Einstein's equations that can not be found using the Schwarzschild like coordinates. 

If we consider that $\omega=(\tilde{A}(r))^2$, the eqs (\ref{4.2})-(\ref{4.4}) reads
\begin{eqnarray}
8\pi T_0^0 & = &  -3(\tilde{A}')^2+2\tilde{A}\tilde{A}'' +\frac{4}{r}\tilde{A}\tilde{A}', \label{4.6} \\
-8\pi T_1^1 & = & (\tilde{A}')^2-\tilde{A}\tilde{A}'\nu' +\nu' \frac{\tilde{A}^2}{r}-\frac{2\tilde{A}\tilde{A}'}{r}, \label{4.7} \\
-8\pi T_2^2 & = & (\tilde{A}')^2-\tilde{A}\tilde{A}'' +\left(\frac{\nu''}{2}+\frac{(\nu')^2}{4}+\frac{\nu'}{2r}\right)\tilde{A}^2 \nonumber \\ & - &\frac{\tilde{A}\tilde{A}'}{r}. \label{4.8}
\end{eqnarray}

At this point is easy to see that the system (\ref{4.2})-(\ref{4.4}) (or (\ref{4.6})-(\ref{4.8})) will not be decoupled if we choose an energy momentum tensor of the form (\ref{3.1}) and consider the particular ansatz for the metric (\ref{3.5}) of the MGD method . Nevertheless, as we will show in next sections, this conditions can be used to obtain new internal analytical solutions (physically acceptable) of Einstein's equations  in isotropic coordinates. 

Finally, in order to avoid the appearance of singular behaviour of the physical variables on the surface of our distribution, we must impose the well known matching conditions between the interior and the exterior space-time geometries. The inner region is defined by the metric (\ref{4.1}) and we will consider that the outer region is described by the vacuum Schwarzschild solution

\begin{equation}
ds^{2}=\left(1-\frac{2M}{r_1}\right) dt^2 -\left(1-\frac{2M}{r_1}\right)^{-1}dr_1^2+r_1^2d\Omega^2,
\label{swis}
\end{equation}
where M denote the total mass of the distributions. 

Now, using (\ref{4.1}) and (\ref{swis}) the matching condition takes the following form  

\begin{eqnarray}
e^{\nu_{\Sigma}} & = & \left(1-\frac{2M}{r_{1\Sigma}}\right), \\
\frac{r_\Sigma}{2}\left(\frac{2}{r_\Sigma}-\frac{w'}{w}\right)_\Sigma \nonumber  & = & \left(1-\frac{2M}{r_{1\Sigma}}\right)^{1/2}, \\
P_r(r_\Sigma) & = & 0,
\end{eqnarray}
were the subscript $\Sigma$ indicates that the quantity is evaluated at the boundary of the distribution. Then its possible to obtain an expression form the total mass of the distribution given by

\begin{equation}
    M = \frac{r_\Sigma}{2w^{1/2}}\left[1-\frac{r^2_\Sigma}{4}\left(\frac{2}{r_\Sigma}-\frac{w'}{w}\right)^2\right]_\Sigma
\end{equation}

\section{Physical acceptability conditions}
\label{sec:4}

Now, solving Einstein's equations does not ensure that the solution will describe any physical system. Indeed, among all the known solutions of Einstein's equations, only a part of them fulfill the physically acceptable conditions (see for example \cite{Delgaty}). 

Then, in order to ensure that the solutions of Einstein's equations are physically acceptable, they must satisfy the following conditions  
\begin{itemize}
\item $P_r$, $P_t$ and $\rho$ are positive and finite inside the distribution.
\item $\frac{dP_r}{dr}$, $\frac{dP_t}{dr}$ and $\frac{d\rho}{dr}$ are monotonically decreasing.
\item Dominant energy condition: $\frac{P_r}{\rho}\leq 1$ \hspace{0.1cm}, \hspace{0.1cm} $\frac{P_t}{\rho}$ $\leq 1$.
\item Causality condition: $0<\frac{dP_r}{d\rho}<1$\hspace{0.1cm}, \hspace{0.1cm}$0<\frac{dP_t}{d\rho}<1$.
\item  The local anisotropy of the distribution should be zero at the center and increasing towards the surface.
\end{itemize}

\section{The algorithms for obtain new solutions}
\label{sec:5}

In this section we present two possibles and inequivalent algorithms to find new analytical anisotropic solutions of Einstein's equations in isotropic coordinates based in the conditions (\ref{3.1}) and (\ref{3.5}) of the MGD method.

\subsection{The first algorithm}
\label{subsec:1}

In order to solve Einstein's equations we  will consider an energy-momentum tensor of the form (\ref{3.1}). In this case, as the geometry remains untouched, Einstein's equations reads as before
\begin{eqnarray}
8\pi \tilde{\rho}(r)&=&\omega''-\frac{5}{4}\frac{\omega'^2}{\omega}+\frac{2}{r}\omega' , \label{5.1} \\
8\pi \tilde{p}_r(r)&=&\frac{1}{4}\frac{\omega'^2}{\omega}-\frac{1}{2}\omega'\nu'+\frac{\nu'\omega-\omega'}{r} , \label{5.2}  \\
8\pi \tilde{p}_t(r)&=&\frac{1}{2}\left( \frac{\omega'^2}{\omega}-\omega''\right)+\left( \frac{\nu''}{2}+\frac{\nu'^2}{4}+\frac{\nu'}{2r}\right)\omega \nonumber \\ & - & \frac{\omega'}{2r}   ,  \label{5.3}
\end{eqnarray}
but with an effective radial pressure $\tilde{p}_r$, effective tangential pressure $\tilde{p}_t$ and effective density $\tilde{\rho}$ that are given by

\begin{equation}
\tilde{\rho}\equiv \rho +\alpha \theta_0^0, \quad \tilde{p}_r \equiv p_r-\alpha\theta_0^0, \quad \tilde{p}_t \equiv p_t -\alpha \theta_2^2. \label{5.4}
\end{equation}
At this point it is easy to see that we have an anisotropic system with three independent equations and seven unknown functions, the same 4 as the isotropic perfect fluid case (\ref{3.6})-(\ref{3.8}) plus $\theta_0^0$ , $\theta_1^1$ , $\theta_2^2$. 

Now, as we mention before, if we try to consider the deformation
\begin{eqnarray}\label{5.5}
\omega\mapsto\tilde{\omega}=\omega+\alpha f,
\end{eqnarray}
is not possible to decouple the system of equations (\ref{5.1})-(\ref{5.3}) due to the presence of $\alpha^2$ terms in $\omega'^2$ and $\omega''$.

However, if we consider a specific combination of equations (\ref{5.1})-(\ref{5.3}) given by
\begin{eqnarray}
8\pi(\tilde{p}_r+\tilde{\rho} +2\tilde{p}_t) &=& \frac{1}{2}\nu'^2\omega +\nu''\omega +\frac{2\nu'\omega}{r}-\frac{1}{2}\omega'\nu'  , \label{5.6}
\end{eqnarray}
and we introduce the minimal geometric deformation
\begin{eqnarray}\label{5.7}
\omega\mapsto\tilde{\omega}=\omega+\alpha f,
\end{eqnarray}
we obtained that this equation splits in two: one equation associated to a perfect fluid and another one related to the gravitational source
\begin{eqnarray}
\hspace{-1cm} 8\pi(p_r+\rho +2 p_t) &=& \frac{1}{2}\nu'^2\omega +\nu''\omega +\frac{2\nu'\omega}{r}  -  \frac{1}{2}\omega'\nu' \label{5.8}\\
\hspace{-1cm} 8\pi(-\theta_1^1+\theta_0^0-2\theta_2^2)&=& \frac{1}{4}\left\lbrace 2\nu'^2f +4\nu''f  \right. \nonumber \\ & + & \left. \frac{8\nu'f}{r} - 2f'\nu' \right\rbrace \label{5.9},  
\end{eqnarray}
respectively.

If we now impose a linear constraint of the form 
\begin{eqnarray}
\theta_0^0-\theta_1^1-2\theta_2^2=-\frac{F(r)}{8\pi}, \label{5.10}
\end{eqnarray}
we have from (\ref{5.9}) that
\begin{equation}
f=e^\nu \nu'^2 r^4 \left(2 \int \frac{F(r)e^{-\nu}}{\nu'^3 r^4}dr +C\right) , \label{5.11-4}
\end{equation}
where $C$ is a constant. Although it is not possible to decouple the system of equations (\ref{5.1})-(\ref{5.3}) after we introduce the deformation, is remarkable that knowing $f$, $f'$ and $f''$, we can obtain $\theta_0^0$,$\theta_1^1$ and $\theta_2^2$ for each value of $\alpha$, from equations (\ref{5.1}), (\ref{5.2}) and (\ref{5.3}) respectively. In fact

%\begin{eqnarray}
%\hspace{-0.9cm}8\pi\theta_0^0 &=& \frac{1}{\omega+\alpha f}\left\lbrace\left[ \omega''+\frac{2}{r}\omega'-8\pi\rho \right]f+\left[\frac{2}{r}\omega - \frac{5}{2}\omega' \right]f' + f''\omega + \alpha\left[\frac{2}{r}f'f-\frac{5}{4}f'^2+ff'' \right] \right\rbrace \\
%\hspace{-0.9cm}8\pi\theta_1^1 &=& \frac{1}{(\omega+\alpha f)}\left\lbrace \left[8\pi P_r -\frac{2}{r}\nu'\omega + \frac{1}{2}\omega'\nu' + \frac{\omega'}{r} \right]f - \left[\frac{1}{2}\omega'-\frac{1}{2}\omega\nu'-\frac{1}{r}\omega \right]f' \right. \nonumber \\
%&-& \left.\alpha\left[\frac{1}{4}{f'}^2-\frac{1}{2}\nu' f'f+\frac{\nu' f^2-f'f}{r} \right]\right\rbrace \\
%\hspace{-0.9cm}8\pi\theta_2^2 &=& \frac{1}{(\omega+\alpha f)}\left\lbrace \left[ 8\pi P_t + \frac{1}{2}\omega'' - \nu''\omega - \frac{1}{2}\nu'^2\omega - \frac{\nu'}{r}\omega + \frac{\omega'}{2r} \right]f - \left[ \omega' - \frac{\omega}{2r}\right]f' +\frac{1}{2}\omega f'' \right. , \nonumber \\
%&-& \left. \alpha\left[ \left(\frac{1}{2}\nu'' + \frac{1}{4}\nu'^2 + \frac{\nu'}{2r} \right)f^2 + \frac{1}{2}f'^2 - \frac{1}{2}f''f-\frac{1}{2r}f'f \right]\right\rbrace
%\end{eqnarray}

\begin{eqnarray}
8\pi\theta_0^0 &=& \frac{f}{\omega+\alpha f}\left\lbrace \frac{5}{4}\frac{\omega'^2}{\omega}+\frac{2}{r}\omega \widetilde{H}_1-\frac{5}{2}\omega' \widetilde{H}_1+\omega \widetilde{H}_2 \right. \nonumber \\ & + & \left. \alpha f\left[\frac{2}{r}\widetilde{H}_1-\frac{5}{4}\widetilde{H}_1^2+\widetilde{H}_2 \right] \right\rbrace, \label{5.12} \\
8\pi\theta_1^1 &=& \frac{f}{(\omega+\alpha f)}\left\lbrace \frac{1}{4}\frac{\omega'^2}{\omega} -\frac{1}{r}\omega\nu' \right. \nonumber \\ & + & \left( \frac{1}{2}\omega\nu'-\frac{1}{2}\omega'+\frac{1}{r}\omega \right)\widetilde{H}_1 \nonumber \\
&-& \left.\alpha f\left[\frac{1}{4}\widetilde{H}_1^2-\left(\frac{1}{2}\nu'-\frac{1}{r}\right) \widetilde{H}_1+\frac{1}{r}\nu' \right]\right\rbrace, \label{5.13} \\
8\pi\theta_2^2 &=&  \frac{f}{(\omega+\alpha f)}\left\lbrace \frac{1}{2}\frac{\omega'^2}{\omega}-\frac{1}{2}\nu''\omega-\frac{1}{4}\nu'^2\omega  \right. \nonumber \\ & - & \frac{1}{2r}\omega\nu'+ \frac{1}{2}\omega \widetilde{H}_2 -\omega' \widetilde{H}_1+\frac{1}{2r}\omega \widetilde{H}_1
 \nonumber \\
&-& \alpha f\left[\frac{1}{2}\widetilde{H}_1^2-\frac{1}{2}\widetilde{H}_2+\frac{1}{2}\nu''+\frac{1}{4}\nu'^2+\frac{1}{2r}\nu' \right. \nonumber \\ & - & \left. \left. \frac{1}{2r}\widetilde{H}_1 \right]\right\rbrace, \label{5.14}
\end{eqnarray}  
where we have used eqs (\ref{4.2})-(\ref{4.4}), and 
\begin{eqnarray}
f'&=& \widetilde{H}_1f , \label{5.15}\\
f'' &=& \widetilde{H}_2f, \label{5.16}
\end{eqnarray} 
with
\begin{eqnarray}
\widetilde{H}_1&=&H_1+\frac{2F(r)}{\nu'f}, \\
\widetilde{H}_2 &=& H_2 +\frac{2}{\nu'f}\left[H_1-\frac{\nu''}{\nu'} \right]F(r)+\frac{2}{\nu'f}F'(r), \\
H_1 &=& \nu' + 2\frac{\nu''}{\nu'}+\frac{4}{r} , \label{5.17} \\
H_2 &=& H_1^2+H_1'. \label{5.18}
\end{eqnarray}
 Then, knowing $\theta_0^0$, $\theta_1^1$ and $\theta_2^2$, we can write the expressions for the effective pressures and energy density using (\ref{5.4}). This represent an analytical and (in general) anisotropic solution of Einstein's Equations in isotropic coordinates, different from the original one.

\subsection{The second algorithm}
\label{subsec:2}

Starting with the line element in isotropic coordinate given by 
\begin{equation}
ds^2 = e^{\tilde{\nu}(r)}-\frac{1}{\tilde{A}(r)^2}(dr^2+r^2 d\Omega^2),
\end{equation}
and considering that the energy momentum tensor has the form 
\begin{equation}
T_{\mu \nu}=T^{PF}_{\mu \nu}+\alpha \Theta_{\mu \nu}+\alpha^2 H_{\mu \nu}
\end{equation}
where $T^{PF}_{\mu \nu}$ is the energy momentum tensor of a perfect fluid. Then Einstein's equations leads to the system
\begin{eqnarray}
8\pi \tilde{\rho} & = &  -3(\tilde{A}')^2+2\tilde{A}\tilde{A}'' +\frac{4}{r}\tilde{A}\tilde{A}', \\
8\pi \tilde{P}_r & = & (\tilde{A}')^2-\tilde{A}\tilde{A}'\nu' +\nu' \frac{\tilde{A}^2}{r}  - \frac{2\tilde{A}\tilde{A}'}{r}, \\
8\pi \tilde{P}_t & = & (\tilde{A}')^2-\tilde{A}\tilde{A}''  +  \left(\frac{\nu''}{2}+\frac{(\nu')^2}{4}+\frac{\nu'}{2r}\right)\tilde{A}^2 \nonumber \\ & - & \frac{\tilde{A}\tilde{A}'}{r},
\end{eqnarray}
where

\begin{eqnarray}
\tilde{\rho} & \equiv & \rho + \alpha \Theta^0_{0}+ \alpha^2 H^0_{0}, \label{rt} \\
\tilde{P}_r & \equiv &  P - \alpha \Theta^1_{1}- \alpha^2 H^1_{1}, \label{pr} \\
\tilde{P}_t & \equiv & P - \alpha \Theta^2_{2}- \alpha^2 H^2_{2} \label{pt}. 
\end{eqnarray}

Now assuming that the system with $\alpha \not= 0$ is characterized by 
%If we considerer that the inclusion of the sources $\Theta_{\mu \nu}$ and $H_{\mu \nu}$ are encoded in the deformation of the espacial component of the metric
\begin{equation}
\tilde{A}(r)=A(r)+\alpha f(r),
\end{equation}
and supposing that the sets of functions $\{\nu, A, \rho, P \}$ and $\{\nu, f,H^0_{0},H^1_{1},H^2_{2}\}$ are solutions of Einstein's equations we have that the above equation system is decomposed in the following systems: the first
\begin{eqnarray}
8\pi \rho  & = &  -3(A')^2+2AA'' +\frac{4}{r}AA', \\
8\pi P  & = & (A')^2-AA'\nu' +\nu' \frac{A^2}{r}-\frac{2AA'}{r} , \\
8\pi P  & = &  (A')^2-AA'' +\left(\frac{\nu''}{2}+\frac{(\nu')^2}{4}+\frac{\nu'}{2r}\right)A^2 \nonumber \\ & - & \frac{AA'}{r},
\end{eqnarray}
the second 
\begin{eqnarray}
8\pi H^0_{0} & = &  -3(f')^2+2ff'' +\frac{4}{r}ff', \label{h0} \\
-8\pi H^1_{1}  & = & (f')^2-ff'\nu' +\nu' \frac{f^2}{r}-\frac{2ff'}{r}, \label{h1} \\
-8\pi H^2_{2}  & = &  (f')^2-ff'' +\left(\frac{\nu''}{2}+\frac{(\nu')^2}{4}+\frac{\nu'}{2r}\right)f^2 \nonumber \\ & - & \frac{ff'}{r} \label{h2},
\end{eqnarray}
which are both systems of Einstein's equations, and a third one
\begin{eqnarray}
8\pi \Theta^0_{0} & = &  -6A'f'+2(Af''+A''f) \nonumber \\ & + & \frac{4}{r}(Af'+A'f), \label{t0} \\
-8\pi \Theta^1_{1}  & = & 2A'f'-(Af'+A'f)\left(\nu'+\frac{2}{r}\right) \nonumber \\ & + & 2\nu' \frac{fA}{r} , \label{t1} \\
-8\pi \Theta^2_{2}  & = &  2A'f'-(Af''+A''f) \nonumber \\ & + & 2\left(\frac{\nu''}{2}+\frac{(\nu')^2}{4}+\frac{\nu'}{2r}\right)Af \nonumber \\ & - & \frac{(Af'+A'f)}{r} \label{t2},
\end{eqnarray}
which is not. 

Now in order to solve the systems and find the deformation function $f$ we could propose any relation of the components of the sources. In the MGD method the solutions were found imposing some simple mimic constrains. In this case, a simple possibility to solve the system is to assume 
\begin{equation}
\Theta^0_0 - 2\Theta^2_2- \Theta^1_1 = -\frac{\tilde{F}(r)}{8\pi},
\end{equation}
which leads to 
\begin{equation}
f'-f\left(2\frac{\nu''}{\nu}+\nu'+\frac{4}{r}-\frac{A'}{A}\right) = \frac{\tilde{F}}{A\nu'}, \label{dffe}
\end{equation}
whose solution is 
\begin{equation}
f(r)=\frac{r^4 (\nu')^2e^\nu}{A}\left(\int\frac{\tilde{F}(r)}{r^4 (\nu')^3e^\nu}dr+C\right),  
\end{equation}
where $C$ is an integration constant. Then, if we choose an known solution of Einstein's equations in isotropic coordinates is possible to obtain new analytical solutions with local anisotropy in pressures.

\section{New exact anisotropic solutions of Einstein's equations in isotropic coordinates}
\label{sec:6}
\subsection{Using the first method}
\label{subsec:3}
In order to verify the first algorithm, let us choose a known perfect fluid solution in isotropic coordinates as Gold III \cite{Gold} solution
\begin{eqnarray}
e^\nu & = & D \left(\frac{g-1}{g+1} \right), \label{6.40} \\
\frac{1}{\omega} & = & B\left(\frac{g+1}{g}\right)^2, \quad  \label{6.41}  \\
P(r) & = & \frac{b}{2\pi B}\frac{1}{(g+1)^2}\left[ \frac{g}{(g+1)^2}-br^2\right],\label{6.42}  \\
\rho (r)& = & \frac{b}{2\pi B}\frac{1}{(g+1)^2}\left[\frac{3g(g-1)}{(g^2-1)^{1/2}}-br^2(3-2g)\right],\label{6.43}  \\
g(r) & = & \cosh(a+br^2), \label{6.44} 
\end{eqnarray}
where $D$, $B$, $a$ and $b$ are constants. This solution will be regular at the origin only if 
\begin{equation}
B=\frac{(\exp(2a)+1)^2}{(\exp(a)+1)^4}, \label{6.45} 
\end{equation}
nevertheless  this equation will  not be necessary satisfied in the new solutions obtained from this one. 

\subsubsection*{Solution 1}
\label{subsubsec:1}
If we consider $F(r)=0$, for simplicity, it can be check that the deformation function takes the form
\begin{eqnarray}
f=\frac{16CDr^6b^2}{(g+1)^2}, \label{6.46} 
\end{eqnarray}
with $C$ an integration constant. Therefore
\begin{eqnarray}
\theta_1^1 &=& \frac{Bf}{8\pi(g^2+\alpha B f(g+1)^2)}\left[ S_1-\alpha f(g+1)^2\tilde{S}_1\right], \\
\theta_0^0 &=& \frac{Bf}{8\pi(g^2+\alpha B f(g+1)^2)}\left[ S_0+\alpha f(g+1)^2\tilde{S}_0\right], \\
\theta_2^2 &=& \frac{Bf}{8\pi(g^2+\alpha B f(g+1)^2)}\left[ S_2-\alpha f(g+1)^2\tilde{S}_2\right],
\end{eqnarray}
where
\begin{eqnarray}
S_1B &=&\frac{g'^2}{(g+1)^2}-8\frac{b^2rg^2}{g'}+4\frac{b^2r^2g^2}{g'}H_{1} \nonumber \\ & - & \frac{gg'}{(g+1)}H_{1}+\frac{g^2}{r}H_{1}, \\
\tilde{S}_1 &=&\frac{1}{4}H_{1}^2 - 4\frac{b^2r^2}{g'}H_{1} - \frac{1}{r}H_{1}+2\frac{r}{g'}, \\
S_0B &=&5\frac{g'^2}{(g+1)^2}+2\frac{g^2}{r}H_{1} -5\frac{gg'}{(g+1)}H_{1} \nonumber \\ & + &g^2H_2 , \\
\tilde{S}_0 &=& 2\frac{1}{r}H_{1}- \frac{5}{2} \frac{b}{g'}H_{1}^2 + H_2, \\
S_2B &=& 2\frac{g'^2}{(g+1)^2}-4\frac{b^2rg^2}{g'}+16\frac{b^4r^4g^2}{g'^2}(g-1) \nonumber \\
&-& 4\frac{b^2rg^2}{g'}+\frac{1}{2}g^2H_2 -\frac{gg'}{(g+1)}H_{1}+\frac{1}{2}\frac{g^2}{r}H_{1}, \\
\tilde{S}_2 &=&\frac{b}{g'}H_{1}^2-\frac{1}{2}H_2 +8\frac{b^2r}{g'}- 16\frac{b^4r^4}{g'^2}(g-1) \nonumber \\ & - &  \frac{1}{2r}H_{1},
\end{eqnarray}
and
\begin{eqnarray}
H_1 &=& \frac{6}{r}+\frac{8b^2r^2}{g'}(1-g), \\
H_2 &=& H_1^2 + H_1' .
\end{eqnarray}

We can write now the effective energy density and pressures as (\ref{5.4}). 

\begin{figure}
\resizebox{0.4\textwidth}{!}{%
  \includegraphics{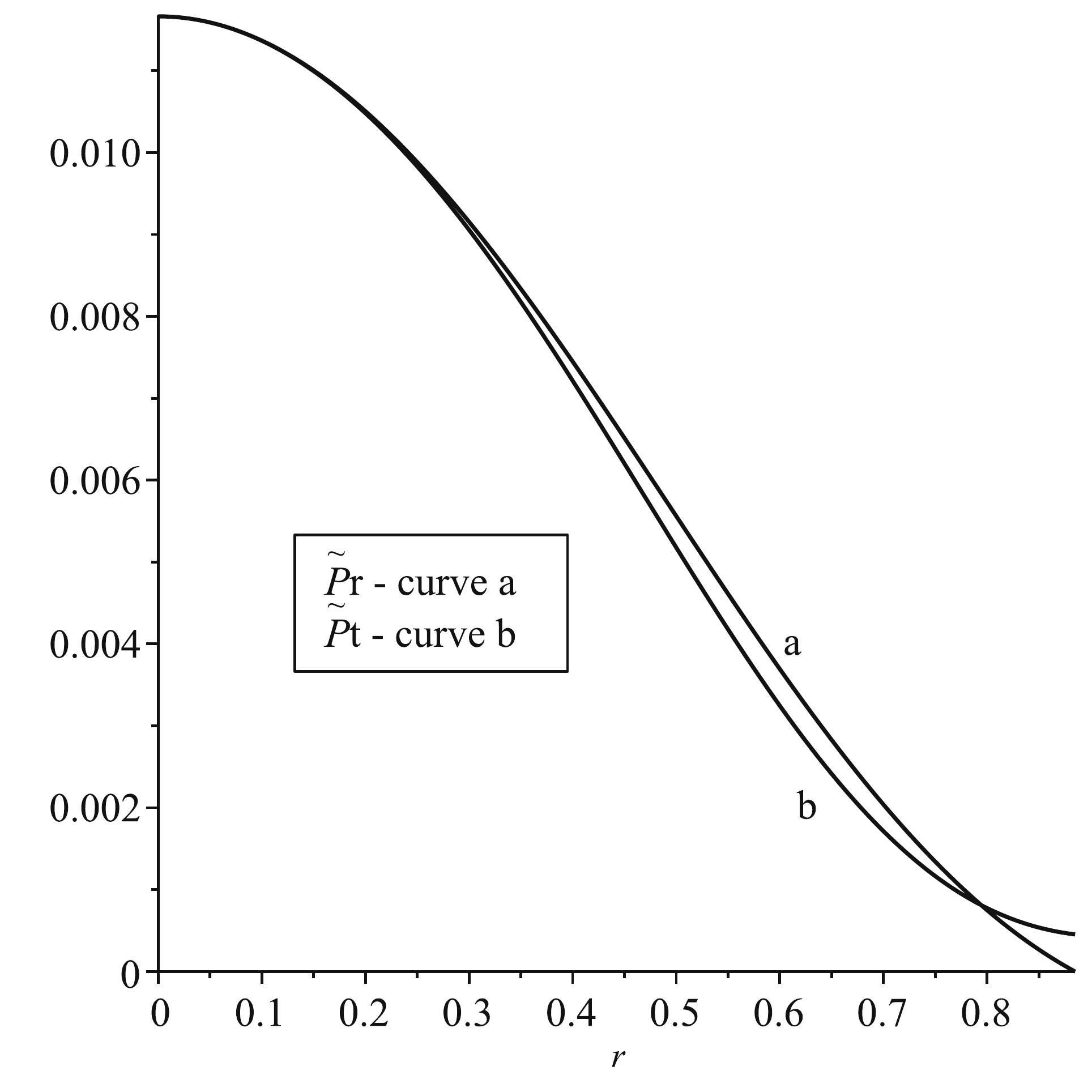}
}
\caption{Pressures for the solution obtained with the first algorithm with $F=0$ from the isotropic Gold III solution.}
\label{fig:1}       % Give a unique label
\end{figure}

\begin{figure}
\resizebox{0.4\textwidth}{!}{%
  \includegraphics{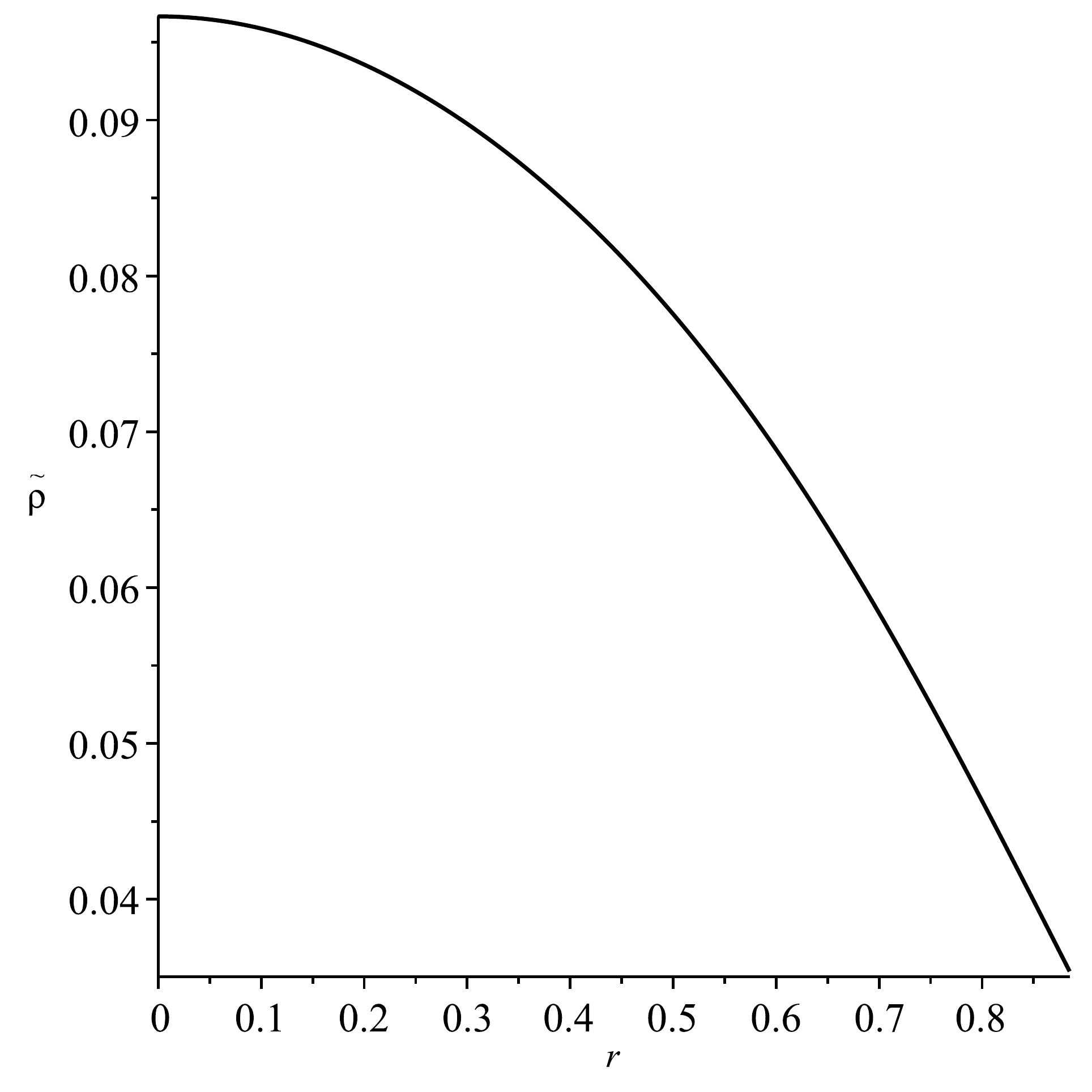}
}
\caption{Energy density for the solution obtained with the first algorithm with $F=0$ from the isotropic Gold III solution.}
\label{fig:2}       % Give a unique label
\end{figure}

\begin{figure}
\resizebox{0.4\textwidth}{!}{%
  \includegraphics{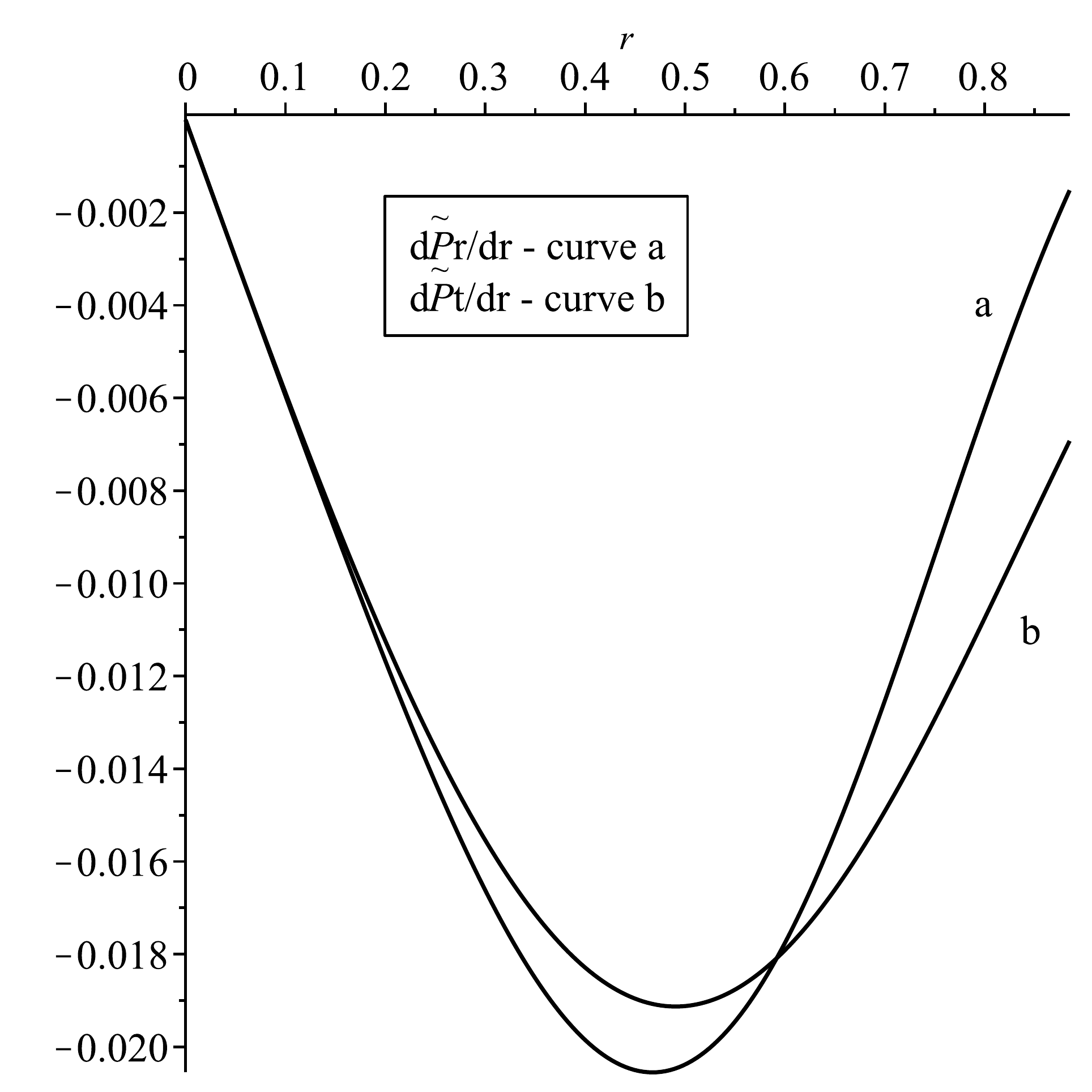}
}
\caption{Gradients of the pressures for the solution obtained with the first algorithm with $F=0$ from the isotropic Gold III  solution.}
\label{fig:3}       % Give a unique label
\end{figure}

\begin{figure}
\resizebox{0.4\textwidth}{!}{%
  \includegraphics{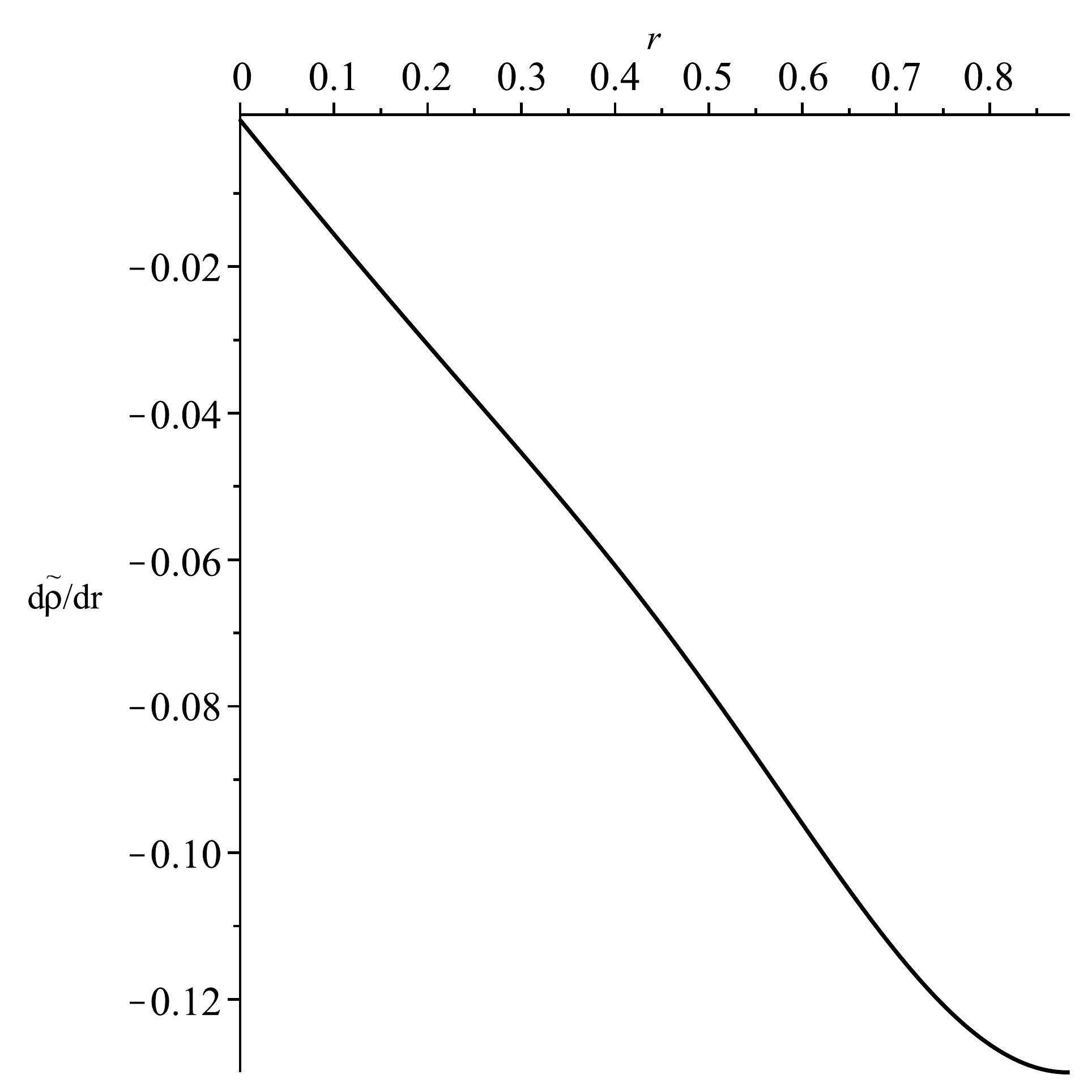}
}
\caption{Gradient of the energy density for the solution obtained with the first algorithm with $F=0$ from the isotropic Gold III solution.}
\label{fig:4}       % Give a unique label
\end{figure}

\begin{figure}
\resizebox{0.4\textwidth}{!}{%
  \includegraphics{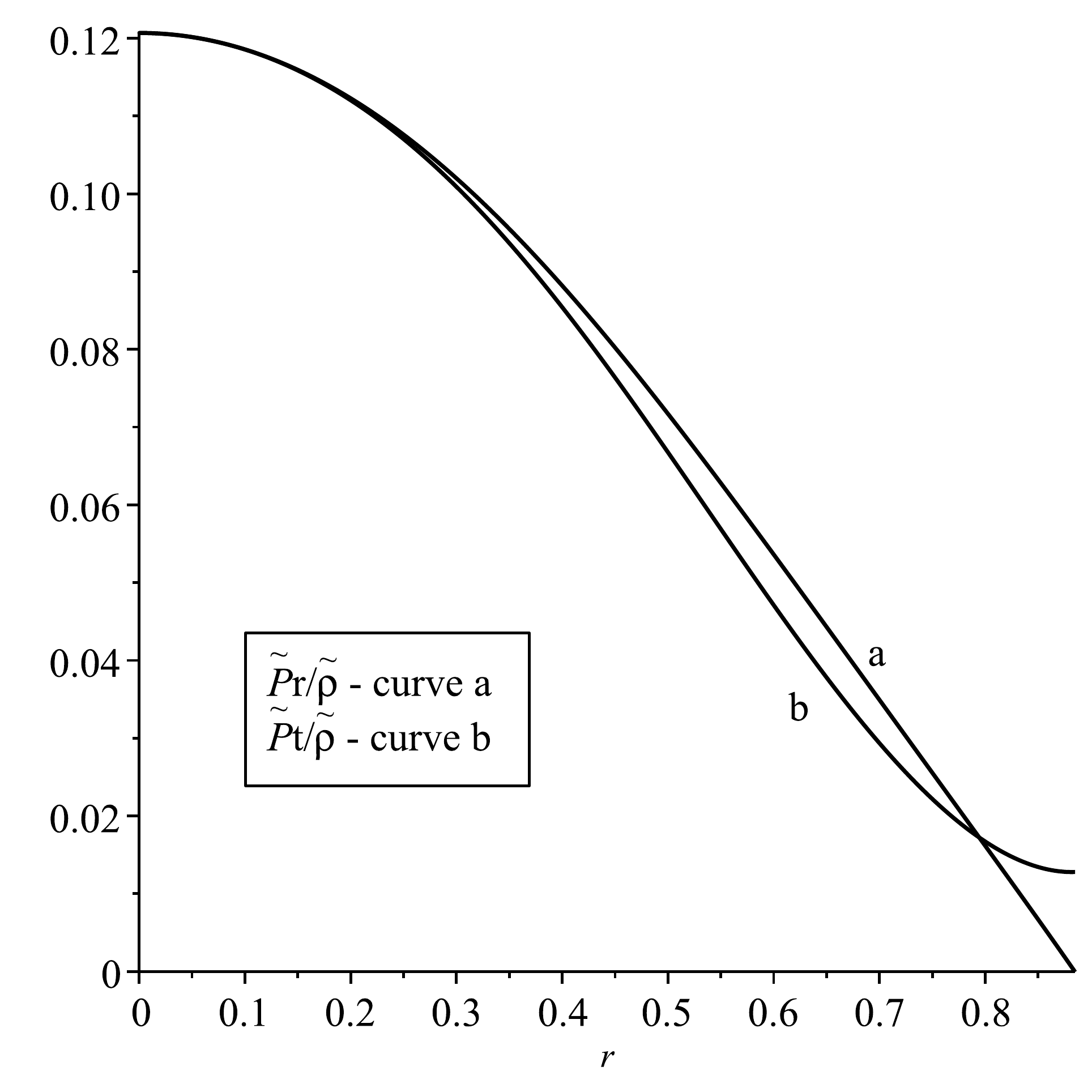}
}
\caption{Dominant energy condition for the solution obtained with the first algorithm with $F=0$ from the isotropic Gold III solution.}
\label{fig:5}       % Give a unique label
\end{figure}

\begin{figure}
\resizebox{0.4\textwidth}{!}{%
  \includegraphics{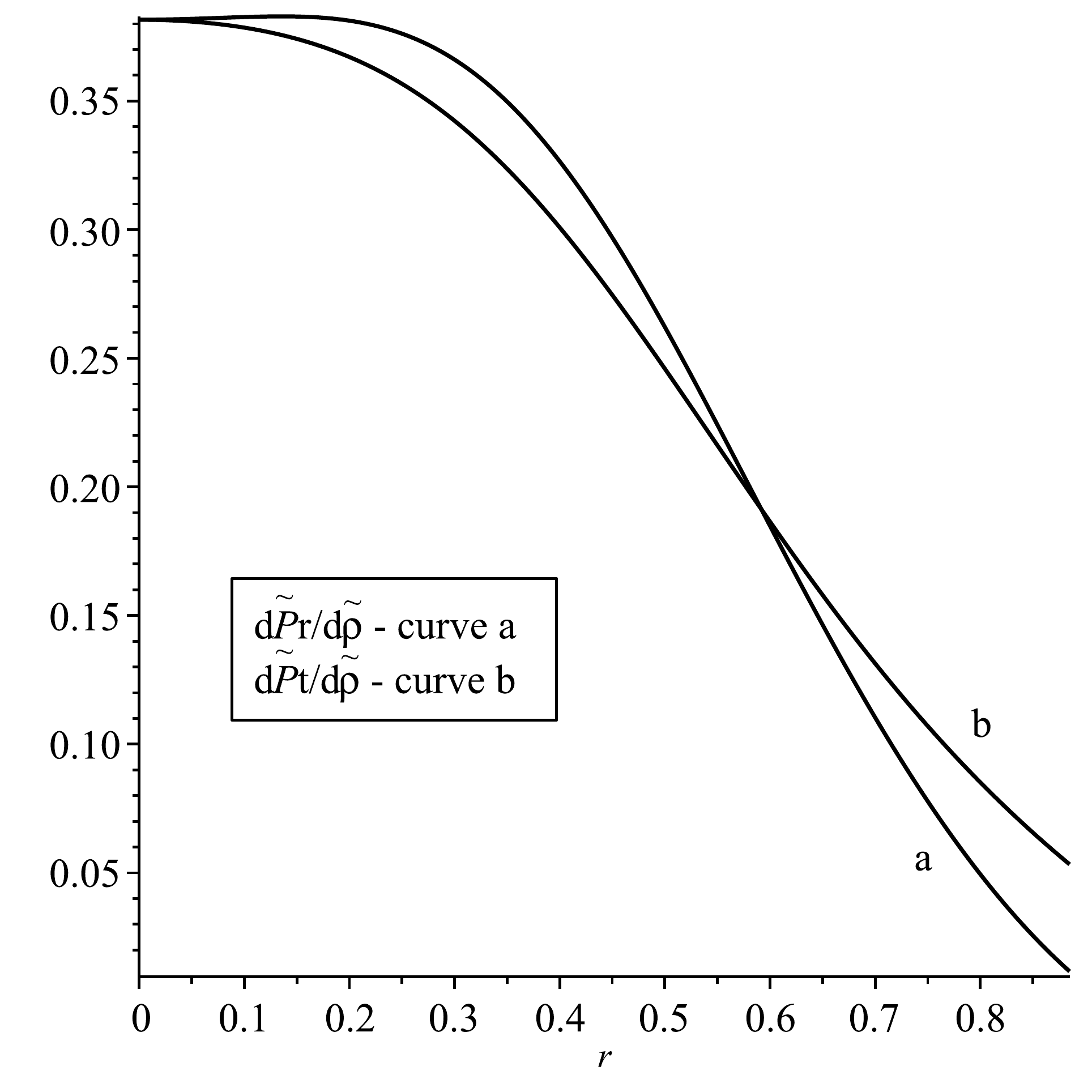}
}
\caption{Causality condition for the solution obtained with the first algorithm with $F=0$ from the isotropic Gold III solution.}
\label{fig:6}       % Give a unique label
\end{figure}

Now, is easy to see that for $a=2$, $b=1$ and $\tilde{\alpha}=D\alpha=0.05$, we found that $r_\Sigma=0.885$, $r_{1\Sigma}=0.7853$, $D=0.881$ and $M=0.126$. The pressures and the energy density are plotted in figures (\ref{fig:1})-(\ref{fig:2}) while the acceptability condition are plotted in figures (\ref{fig:3})-(\ref{fig:6}).

%%%%%%%%%%%%%%%%%%%%%%%%%%%%%%%%Graficos Gold III %%%%%%%%%%%%%%%%%%%%%%%%%%%%%%%%%%%%%%%%%%

%%%%%%%%%%%%%%%%%%%%%%%%%%%%%%%%%%%%%%Narai%%%%%%%%%%%%%%%%%%%%%%%%%%%%%%%%%

\begin{figure}
\resizebox{0.4\textwidth}{!}{%
  \includegraphics{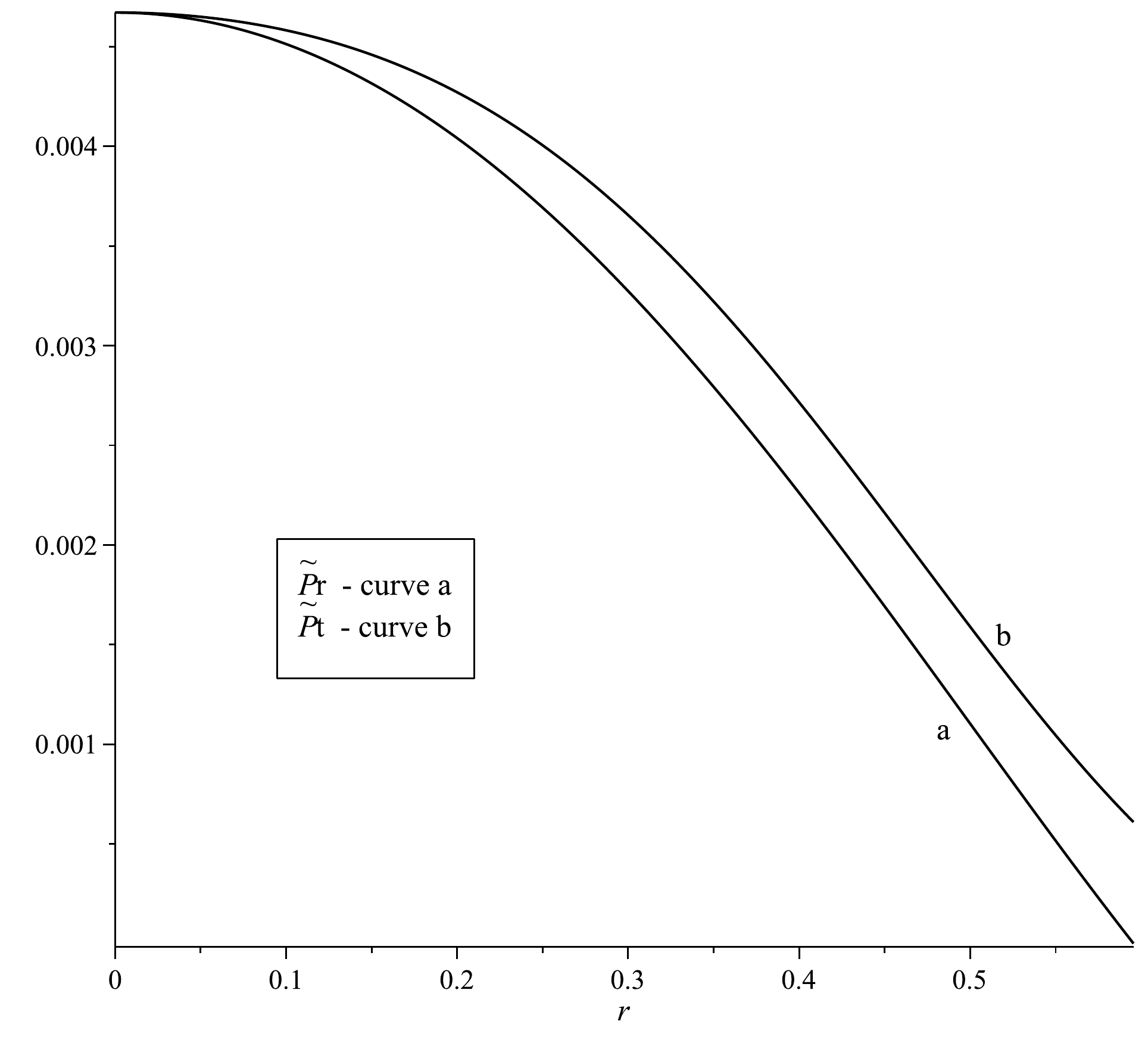}
}
\caption{Pressures for the solution obtained with the first algorithm with $F\not=0$ from the isotropic Gold III solution.}
\label{fig:7}       % Give a unique label
\end{figure}

\begin{figure}
\resizebox{0.4\textwidth}{!}{%
  \includegraphics{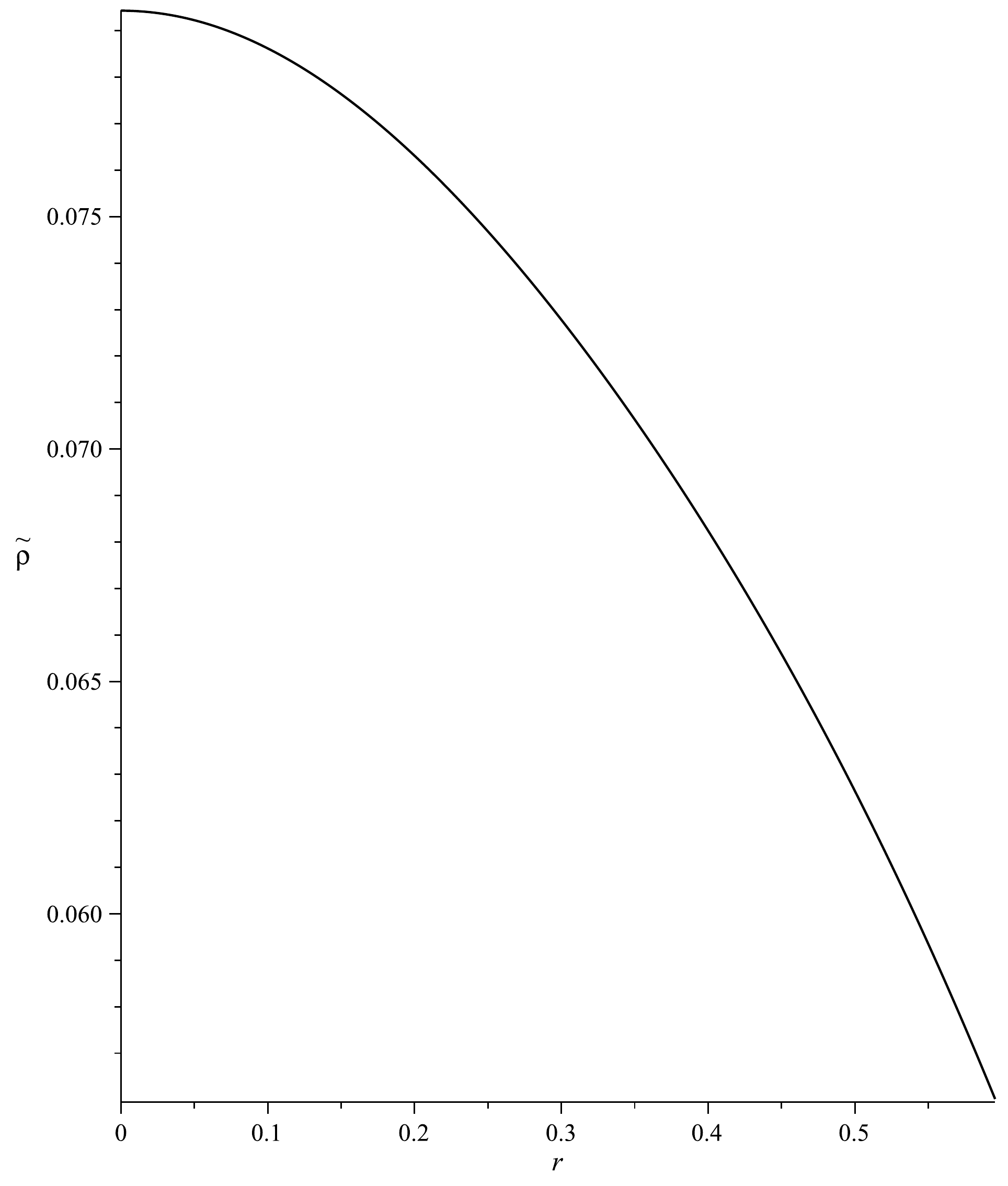}
}
\caption{Energy density for the solution obtained with the first algorithm with $F\not=0$ from the isotropic Gold III solution.}
\label{fig:8}       % Give a unique label
\end{figure}

\begin{figure}
\resizebox{0.4\textwidth}{!}{%
  \includegraphics{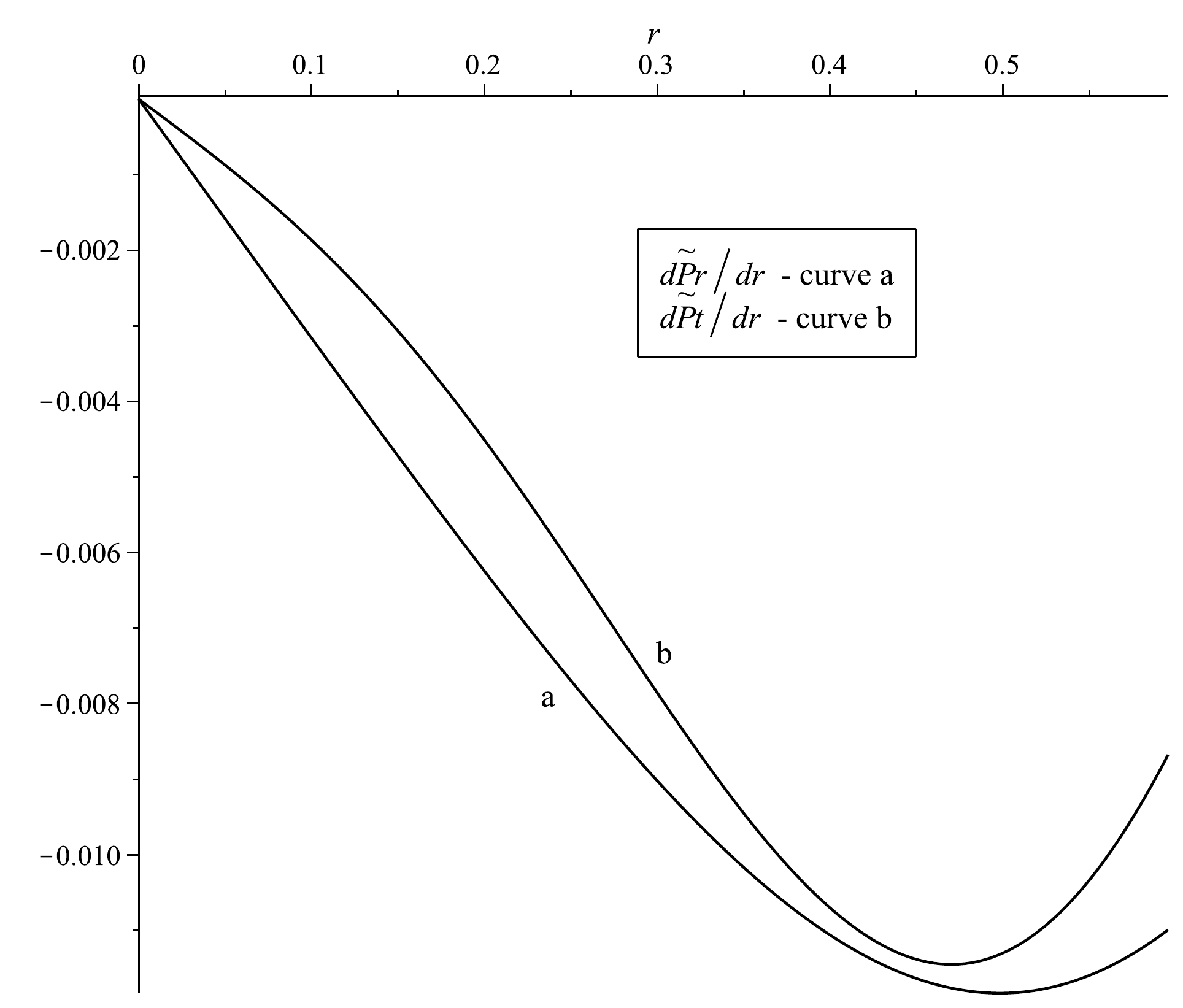}
}
\caption{Gradients of the pressures for the solution obtained with the first algorithm with $F\not=0$ from the isotropic Gold III solution.}
\label{fig:9}       % Give a unique label
\end{figure}

\begin{figure}
\resizebox{0.4\textwidth}{!}{%
  \includegraphics{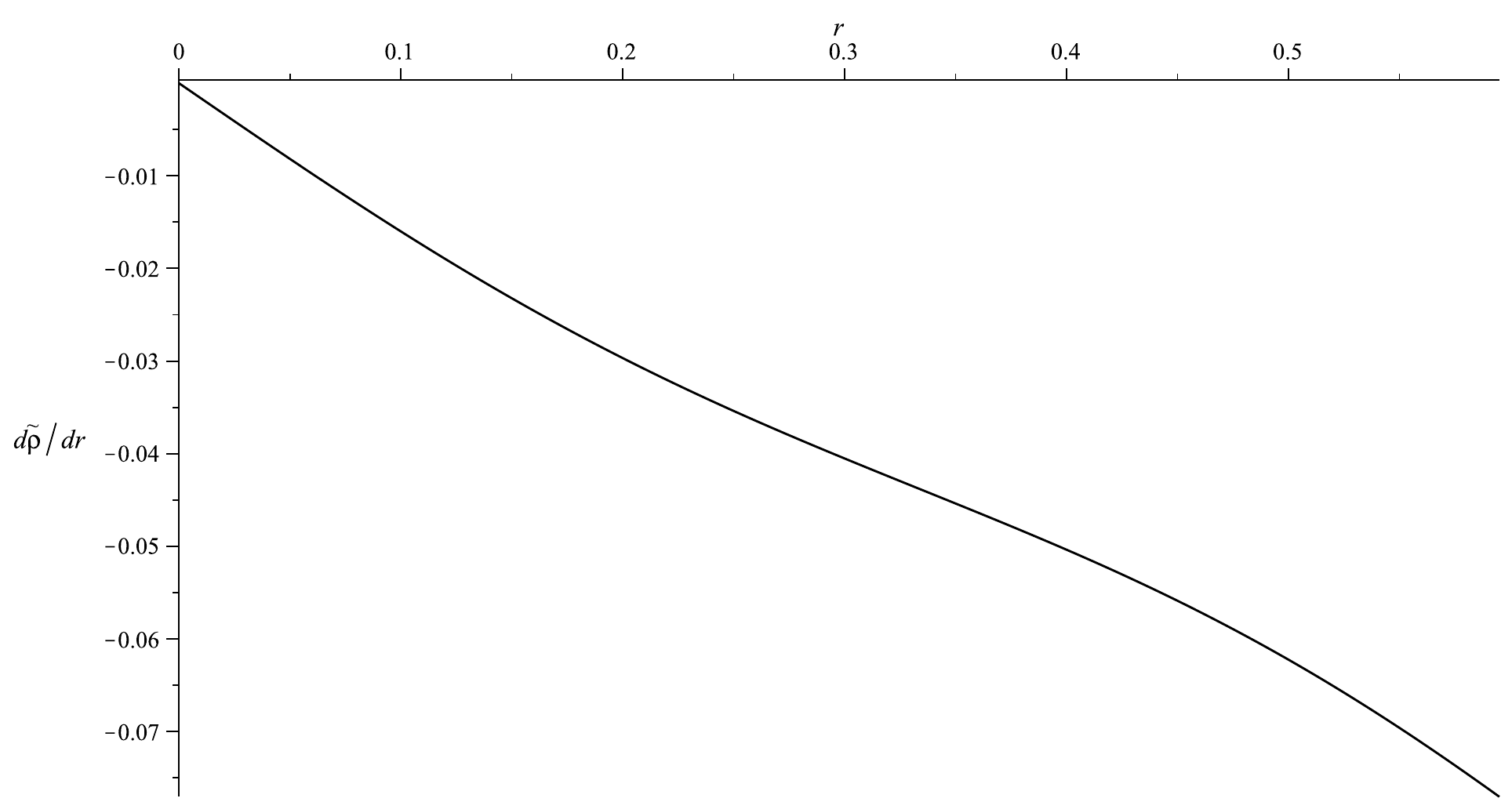}
}
\caption{Gradient of the energy density for the solution obtained with the first algorithm with $F\not=0$ from the isotropic Gold III solution.}
\label{fig:10}       % Give a unique label
\end{figure}

\begin{figure}
\resizebox{0.4\textwidth}{!}{%
  \includegraphics{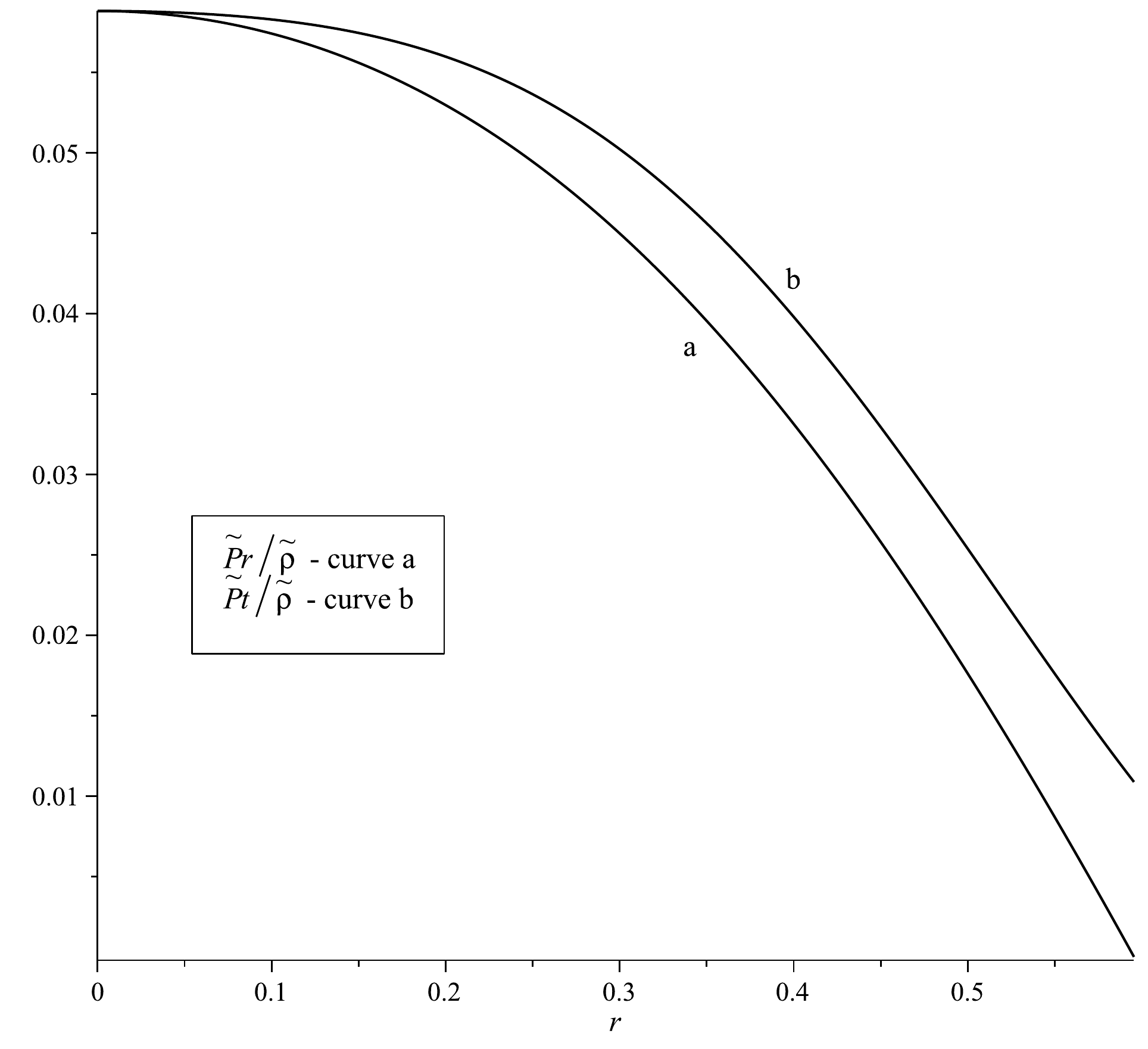}
}
\caption{Dominant energy condition for the solution obtained with the first algorithm with $F\not=0$ from the isotropic Gold III solution.}
\label{fig:11}       % Give a unique label
\end{figure}

\begin{figure}
\resizebox{0.4\textwidth}{!}{%
  \includegraphics{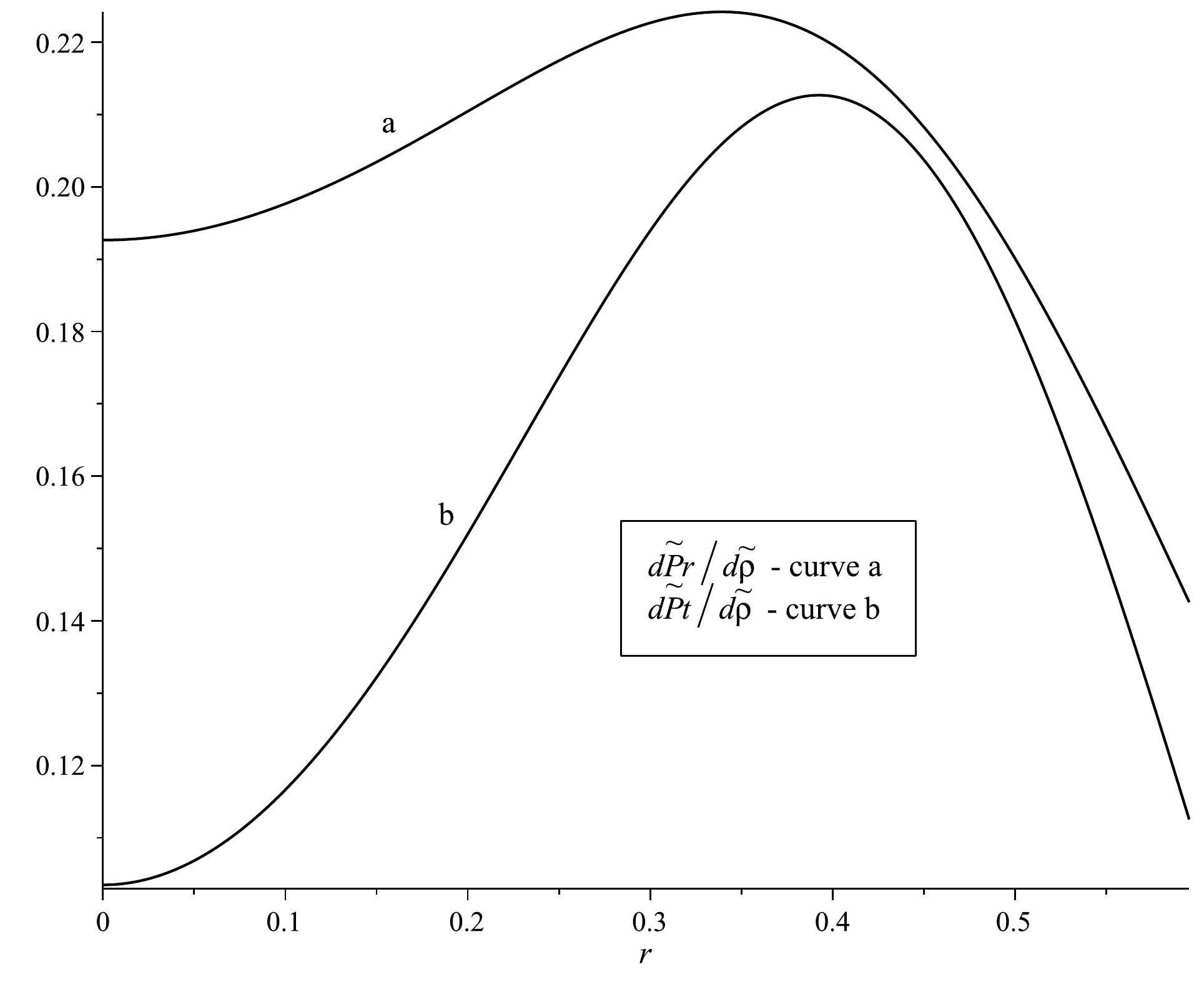}
}
\caption{Causality condition  for the solution obtained with the first algorithm with $F\not=0$ from the isotropic Gold III solution.}
\label{fig:12}       % Give a unique label
\end{figure}

\subsubsection*{Solution 2}
\label{subsubsec:2}

The latter solution can be generalized to the case where $F(r)\neq 0$. In particular, if we choose 
\begin{eqnarray}
F(r)=\frac{16DGb^2\nu'}{r(g+1)^2}
\end{eqnarray}
with $G$ a proportionality constant, then 
\begin{eqnarray}
B=\frac{3}{3+8\alpha Db^2G}\frac{(e^{2a}+1)^2}{(1+e^a)^4}
\end{eqnarray} 
and it can be seen that
\begin{eqnarray}
f=\frac{16Db^2r^6}{(g+1)^2}\left[ \tilde{C}-\frac{G}{3r^6}\right]
\end{eqnarray}
where $\tilde{C}=2K+C$ and $K$ is an integration constant. Is easy to see that it contains the solution 1 ($K=G=0$).

It can be check that
\begin{eqnarray}
\widetilde{H}_1&=& \frac{6}{r}+\frac{8b^2r^2}{g'}(1-g)+\frac{32DGb^2}{rf(g+1)^2} , \\
\widetilde{H}_2 &=&H_1^2 + H_1' , \nonumber \\
&+& \frac{2}{f}\left[ \frac{5}{r}+\frac{8b^2r^2}{g'}(1-g)-\frac{2g'}{g+1}\right]\frac{16DGb^2}{(g+1)^2r}
\end{eqnarray}

Now, from the matching conditions, is easy to see that for$a=2$, $b=1$ and $\tilde{\alpha}=D\alpha=0.1$, we found that $r_\Sigma=0.5955$, $r_{1\Sigma}=0.6527$ , $M=0.07$ and $D=1.1496$. The pressures, energy density and acceptability conditions are plotted in figures (\ref{fig:7})-(\ref{fig:12}).

\subsection{Using the second method}
\label{subsec:4}
In order to check the method, we choose as seed solution the Gold III solution again, which is given by eqs (\ref{6.40})-(\ref{6.44}) where is easy to see that

\begin{eqnarray}
A=\frac{1}{\sqrt{B}}\frac{g}{g+1}.
\label{Ag}
\end{eqnarray}
Now, using these expressions and the differential equation for $f$ (Eq. (\ref{dffe})) we can rewrite the systems (\ref{h0})-(\ref{h2}) and (\ref{t0})-(\ref{t1}) as

\begin{eqnarray}
8\pi H^0_0 & = & f^2\left[G(r)\left(\frac{6}{r}-G(r)\right) -  \frac{24}{r^2} \right. \nonumber \\ & - & \left.  \frac{8b^2r^2(g^2+g+1)}{g^2(g+1)^2}\right]  \nonumber \\ &+& 2f\left[J(r) + 2\hat{F}\left(\frac{1}{r}-G(r)\right)\right]- 3\tilde{F}^2, \label{h0g} \\
-8\pi H^1_1 & = & f^2 \left\lbrace G(r)\left[G(r)-\frac{2}{r}-\frac{4br}{(g^2-1)^{1/2}}\right] \right. \nonumber \\ & + & \left. \frac{4b}{(g^2-1)^{1/2}}\right\rbrace + \hat{F}^2 \nonumber \\ & + & \hat{F}\left[f\left(2\left[G(r)-\frac{1}{r}\right]-\frac{4br}{\sqrt{g^2-1}}\right)\right], \label{h1g} \\
-8\pi H^2_2 & = & f^2 \left\lbrace \frac{12}{r}+\frac{4b^2r^2(g^2+g+1)}{g^2(g+1)^2} \right. \nonumber \\ & + & G(r)\left[\frac{br}{(g^2-1)^{1/2}}-\frac{2}{r} \right] \nonumber \\ & + & \left. \frac{2br}{(g^2-1)^{1/2}}\left(\frac{br(g^2-1)^{1/2}}{g(g+1)}-\frac{1}{r}\right) \right\rbrace \nonumber \\ & + & f\left[\hat{F}\left(G(r)-\frac{1}{r}\right)-J(r)\right]+\hat{F}^2, \label{h2g}
\end{eqnarray}
and

\begin{eqnarray}
8\pi \Theta_0^0 & = & \frac{fg}{\sqrt{B}(g+1)}\left\lbrace 2G(r)\left[ \left(G(r)+\frac{3}{r}\right) \right. \right. \nonumber \\ &-& \left. \frac{6br(g-1)^{1/2}}{g(g+1)}\right]  \nonumber \\ & + &    \frac{4br(g^2-1)^{1/2}}{g(g+1)}\left(\frac{2br(2-g)}{(g^2-1)^{1/2}}+\frac{3}{r}\right) \nonumber \\ & - & \left. 8\left(\frac{3}{r^2}+\frac{b^2r^2(g^2+g+1)}{g^2(g+1)^2}\right) \right\rbrace  \nonumber  \\ & + & \left[\hat{F}\left(G(r)+\frac{2}{r}-\frac{6br(g-1)^{1/2}}{g(g+1)}\right) \right. \nonumber \\ & +&  J(r)\Big], \label{t0g} \\ 
-8\pi \Theta^1_1 & = &  \frac{fg}{\sqrt{B}(g+1)}\left\lbrace G(r)\left[\frac{4br(g^2-1)^{1/2}}{g+1} \right. \right.  \nonumber \\ & - & \left. g\left(\frac{4br}{(g^2-1)^{1/2}}+\frac{2}{r}\right) \right] - \frac{4br(g^2-1)^{1/2}}{g+1}  \nonumber \\ & \times & \left. \left(\frac{2br}{g^2-1)^{1/2}}+\frac{1}{r}\right) + \frac{8bg}{(g^2-1)^{1/2}} \right\rbrace \nonumber \\ & + & \frac{2g\hat{F}}{\sqrt{B}(g+1)}\left[2br\left(\frac{(g^2-1)^{1/2}}{g(g+1)} \right. \right. \nonumber \\  &-& \left. \left. \frac{1}{(g^2+1)^{1/2}}\right)-\frac{2}{r}\right], \label{t1g} \\
-8\pi \Theta^2_2 & = & \frac{16r^6b^2}{g(g+1)^2} \left\lbrace G(r)\left[\frac{4br(g^2-1)^{1/2}}{(g+1)} \right. \right. \nonumber \\ & + & \left. g\left(\frac{2br}{(g^2-1)^{1/2}}-\frac{2}{r}-G(r)\right)\right] \nonumber \\  & - &  \frac{4br(g^2-1)^{1/2}}{g+1}\left(\frac{br(1-g)}{(g^2-1)^{1/2}}+\frac{1}{r}\right) \nonumber \\ & + & \left. \left(\frac{3}{r^2}+\frac{b^2r^2(g^2+g+1)}{g^2(g+1)^2}-\frac{b}{(g^2-1)^{1/2}}\right) \right\rbrace \nonumber \\ &+& \frac{4br(g^2-1)^{1/2}}{g+1} \left[ \hat{F}\left(\frac{4br(g^2-1)^{1/2}}{g(g+1)}\right) \right. \nonumber \\ &-& \left. \frac{1}{r}-G(r)\right]  \label{t2g},
\end{eqnarray}
where
\begin{eqnarray}
G(r) &=& \frac{2br(g+1-2g^2)}{g(g^2-1)^{1/2}}+\frac{6}{r}, \\
\hat{F} &=& \frac{\tilde{F}}{A\nu'}, \\
J(r) &=& \hat{F}'.
\end{eqnarray}

\subsubsection*{Solution 1}
\label{subsubsec:3}
Let us now study the most simple case were $\tilde{F}=0$ in which 
 
\begin{equation}
f=\frac{16r^6DC\sqrt{B}b^2}{g(g+1)},  \label{fg1}
\end{equation} 
where we can take, without losing generality, $C=1$.

\begin{figure}
\resizebox{0.4\textwidth}{!}{%
  \includegraphics{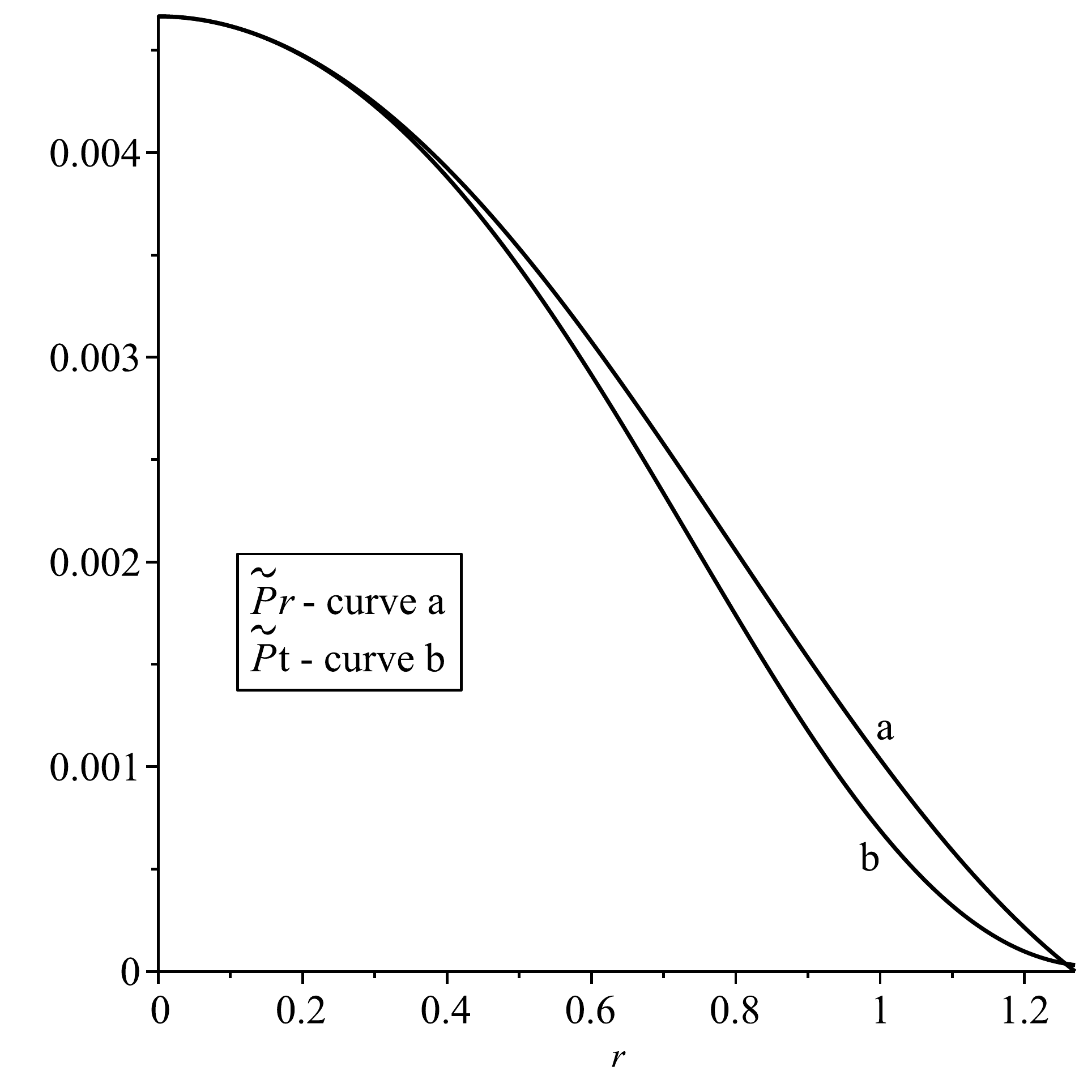}
}
\caption{Pressures for the solution obtained with the second algorithm with $\tilde{F}=0$ from the isotropic Gold III solution.}
\label{fig:13}       % Give a unique label
\end{figure}

\begin{figure}
\resizebox{0.4\textwidth}{!}{%
  \includegraphics{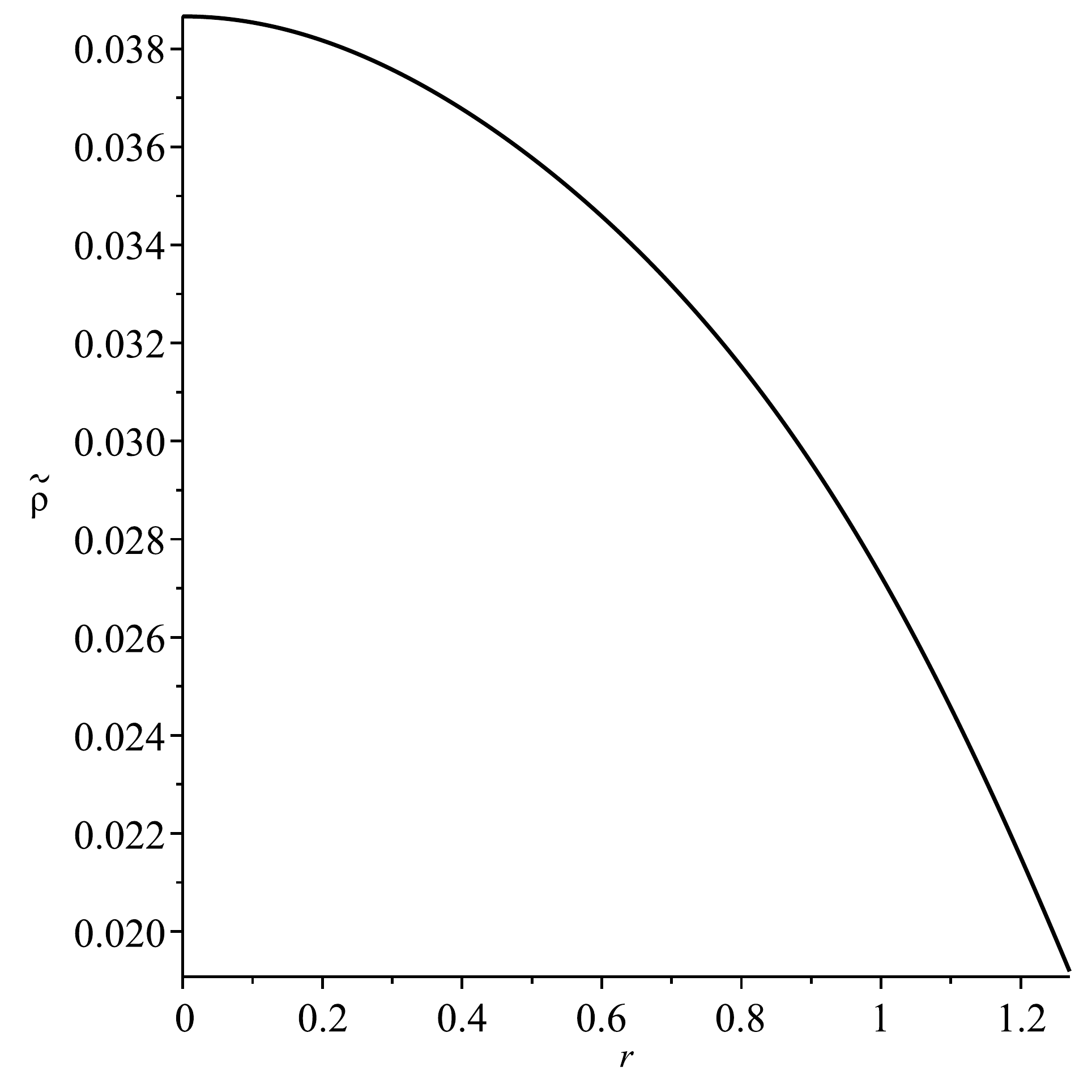}
}
\caption{Energy density for the solution obtained with the second algorithm with $\tilde{F}=0$ from the isotropic Gold III solution.}
\label{fig:14}       % Give a unique label
\end{figure}

\begin{figure}
\resizebox{0.4\textwidth}{!}{%
  \includegraphics{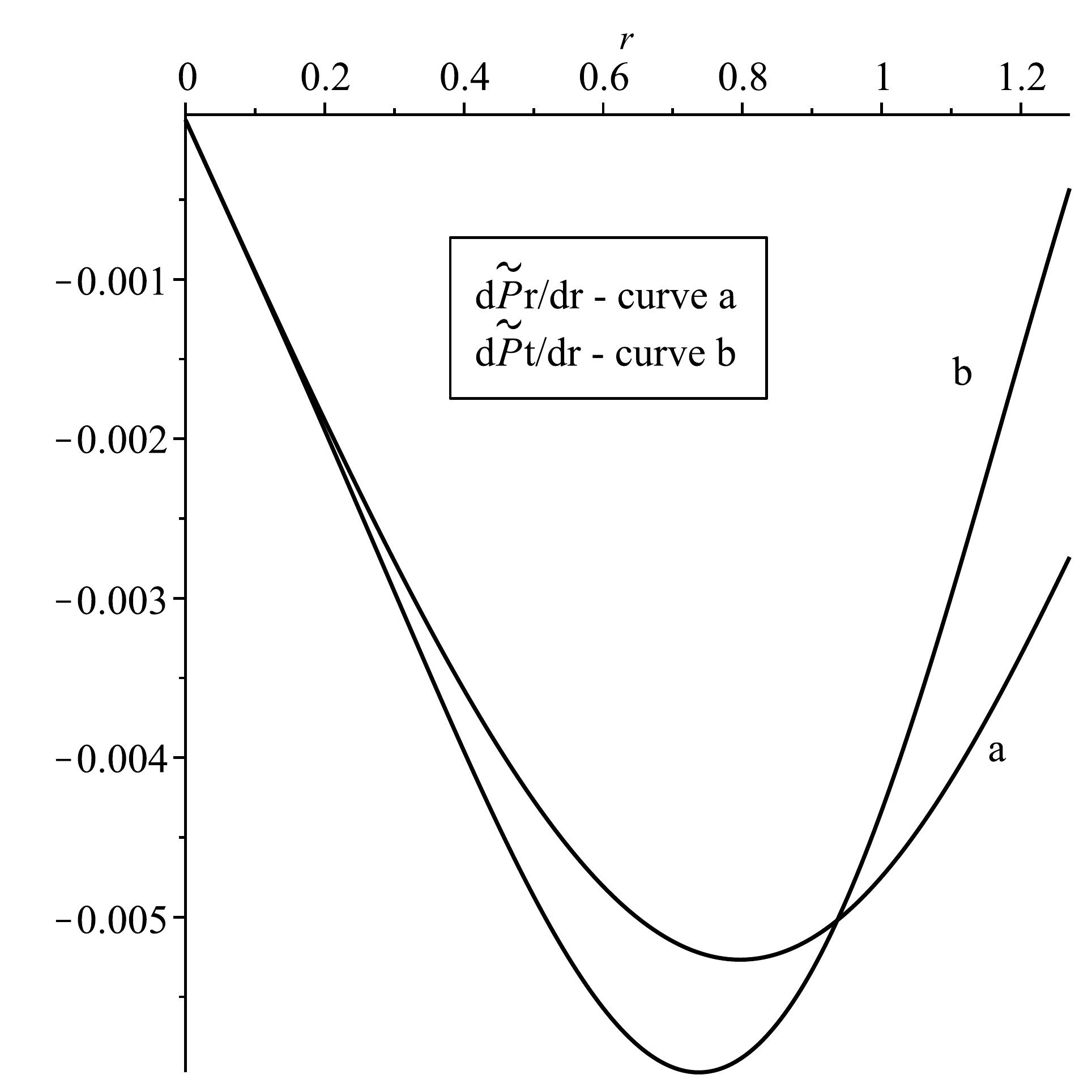}
}
\caption{Gradients of the pressures for the solution obtained with the second algorithm with $\tilde{F}=0$ from the isotropic Gold III solution.}
\label{fig:15}       % Give a unique label
\end{figure}

\begin{figure}
\resizebox{0.4\textwidth}{!}{%
  \includegraphics{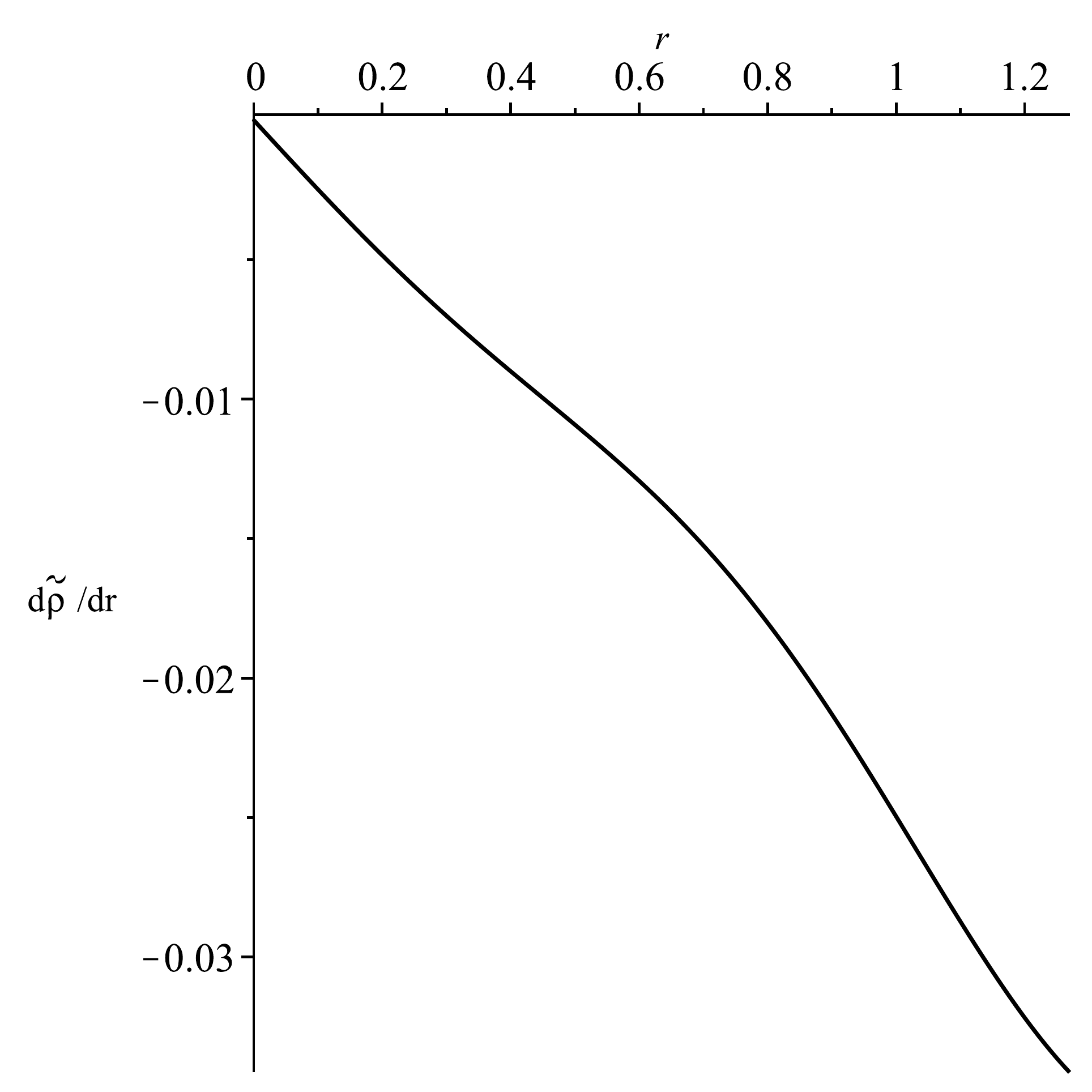}
}
\caption{Gradient of the energy density for the solution obtained with the second algorithm with $\tilde{F}=0$ from the isotropic Gold III solution.}
\label{fig:16}       % Give a unique label
\end{figure}

\begin{figure}
\resizebox{0.4\textwidth}{!}{%
  \includegraphics{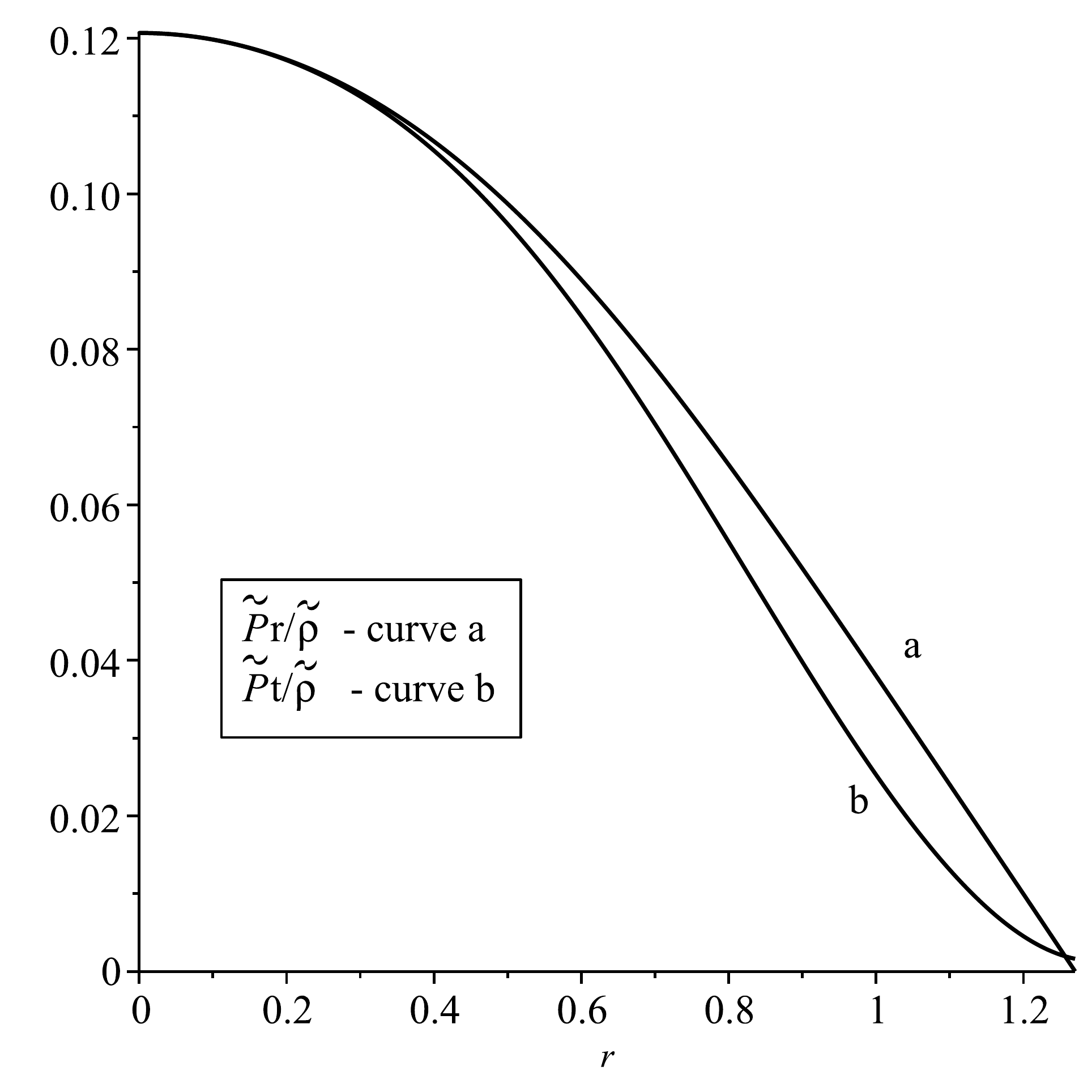}
}
\caption{Dominant Energy condition for the solution obtained with the second algorithm with $\tilde{F}=0$ from the isotropic Gold III solution.}
\label{fig:17}       % Give a unique label
\end{figure}

\begin{figure}
\resizebox{0.4\textwidth}{!}{%
  \includegraphics{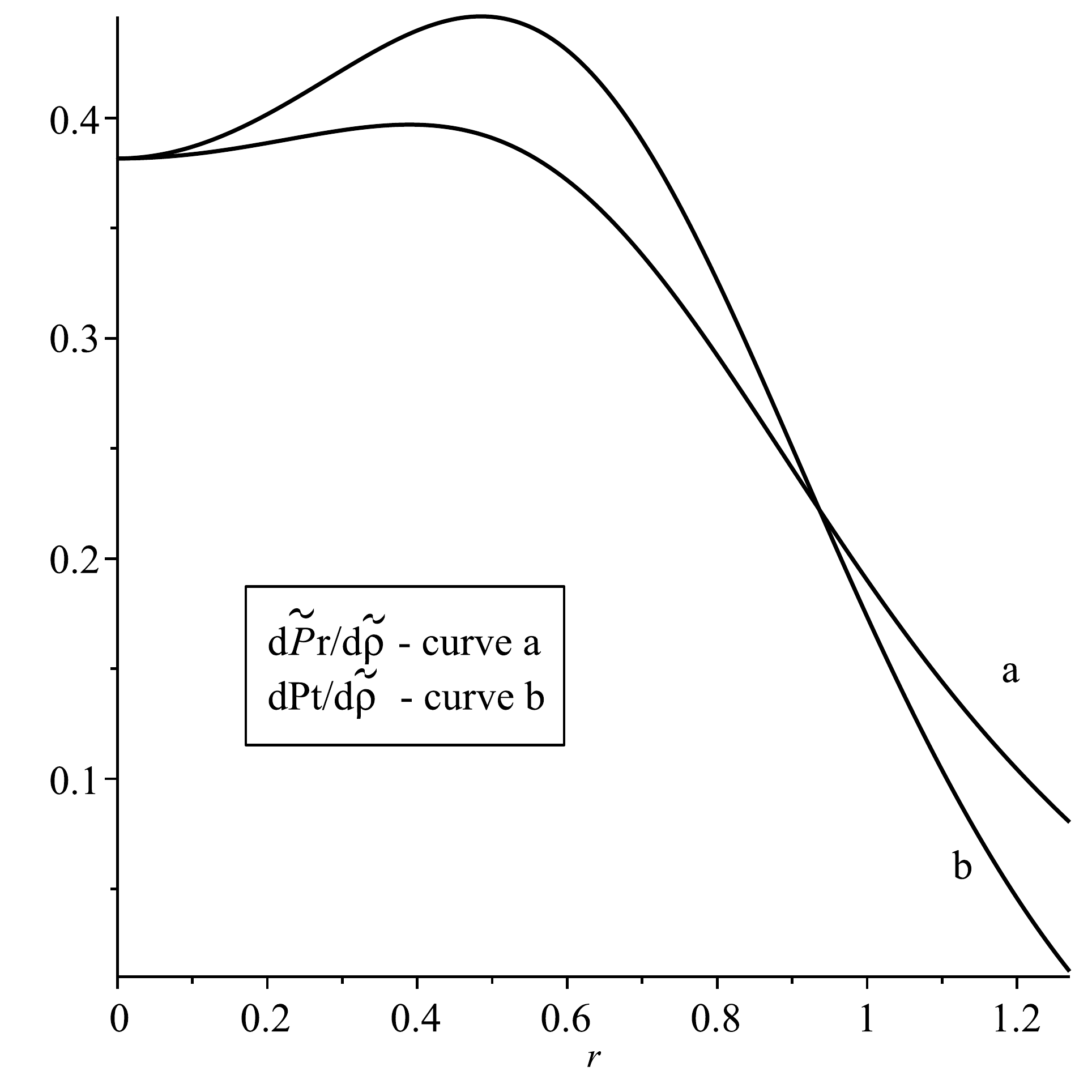}
}
\caption{Causality condition for the solution obtained with the second algorithm with $\tilde{F}=0$ from the isotropic Gold III solution.}
\label{fig:18}       % Give a unique label
\end{figure}

Now, introducing (\ref{fg1}) in the systems (\ref{h0g})-(\ref{h2g}) and (\ref{t0g})-(\ref{t2g}) , we can compute the effective energy density and pressures 
\begin{eqnarray}
\tilde{\rho} & = & \rho +\alpha \Theta^0_0 + \alpha^2 H^0_0, \\
\tilde{P}_r & = & P -\alpha \Theta^1_1 - \alpha^2 H^1_1 ,\\
\tilde{P}_t & = & P -\alpha \Theta^2_2 - \alpha^2 H^2_2. 
\end{eqnarray}
Finally in order to give an example of a physical acceptable solution we can take the following values for the constants, $\alpha D = 0.020$, $a=2$ and $b=0.4$. Then from the matching conditions we obtain that $r_\Sigma=1.270$, $r_{1\Sigma}=1.142$, $M=0.175$, $D=0.922$ and $\alpha=0.022$.The  pressures can be found in the figure (\ref{fig:13}) while the energy density is in the figure (\ref{fig:14}). All the acceptability condition are plotted in the figures (\ref{fig:15})-(\ref{fig:18}).   

%%%%%%%%%%%%%%%%%%%%%%%%%%%%%%%%%%%%Graficos con gold III %%%%%%%%%%%%%%%%%%%%%%%%%%%%%%%%%%%%%%%%%%%%%

%%%%%%%%%%%%%%%%%%%%%%%%%%%%%%%%%%%%%Graficos Narai %%%%%%%%%%%%%%%%%%%%%%%%%%%%%%%%%%%%%

\begin{figure}
\resizebox{0.4\textwidth}{!}{%
  \includegraphics{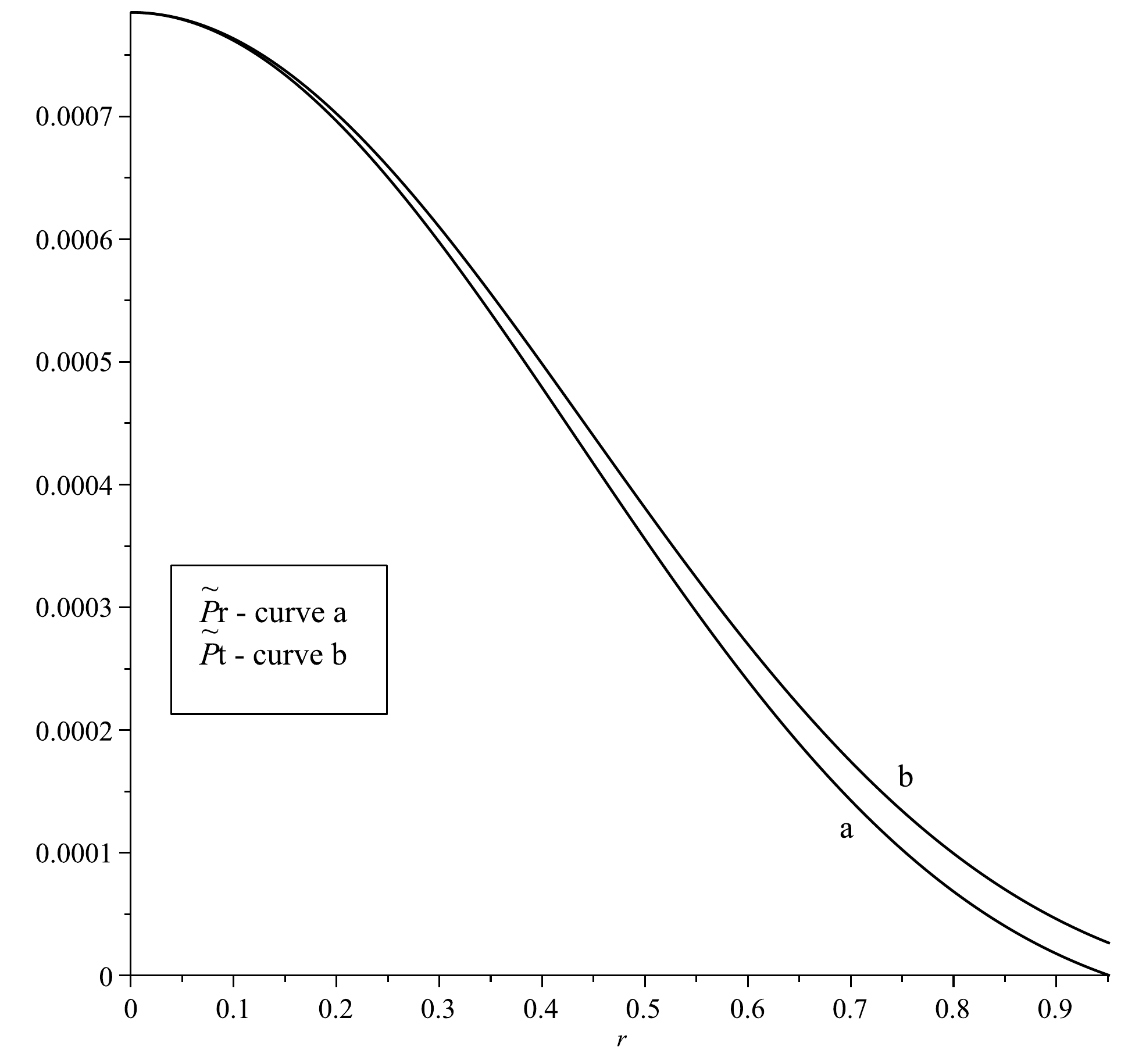}
}
\caption{Pressures for the solution obtained with the second algorithm with $\tilde{F}\not=0$ form the isotropic Gold III solution.}
\label{fig:19}       % Give a unique label
\end{figure}

\begin{figure}
\resizebox{0.4\textwidth}{!}{%
  \includegraphics{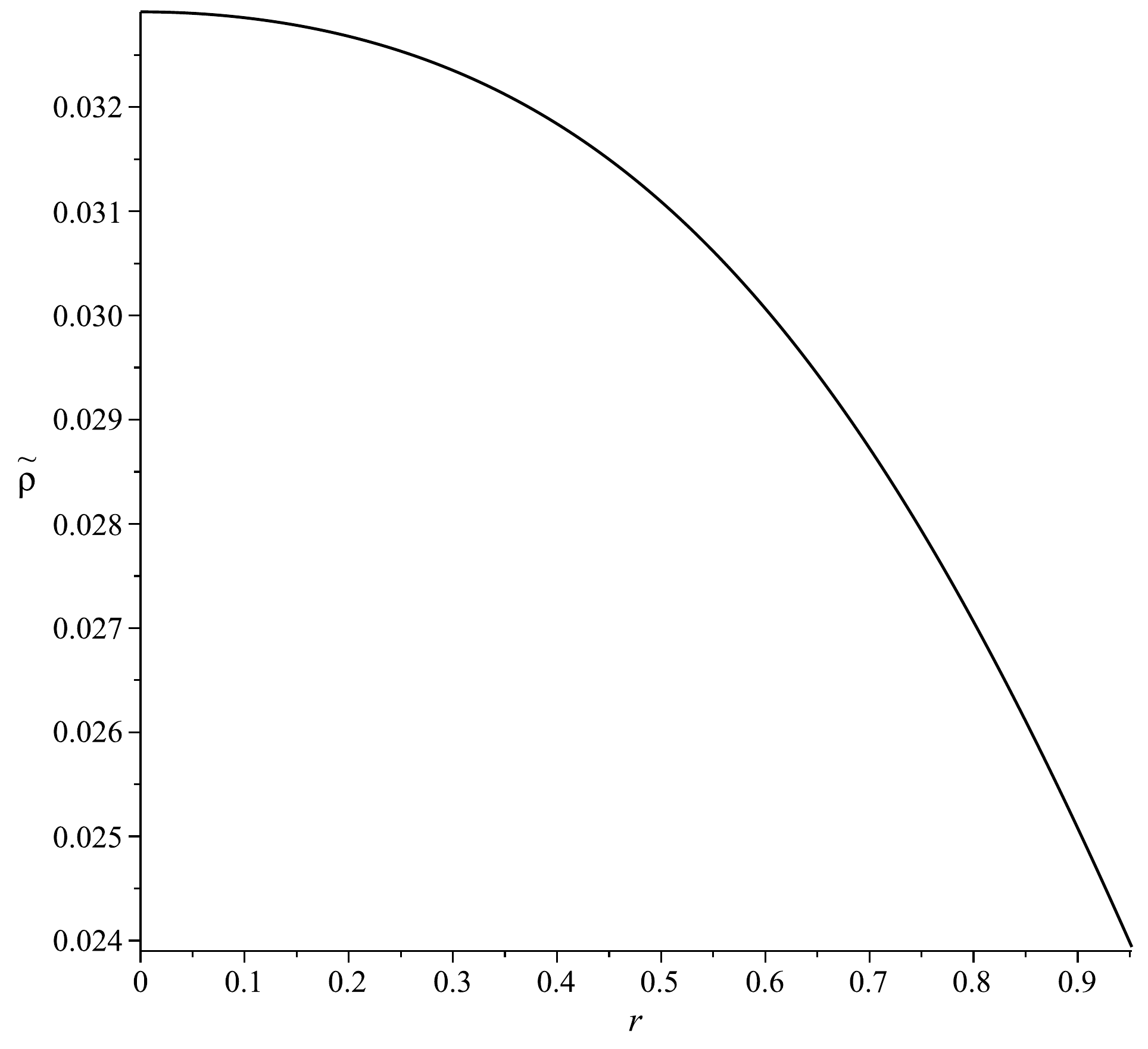}
}
\caption{Energy density for the solution obtained with the second algorithm with $\tilde{F}\not=0$ form the isotropic Gold III solution.}
\label{fig:20}       % Give a unique label
\end{figure}

\begin{figure}
\resizebox{0.4\textwidth}{!}{%
  \includegraphics{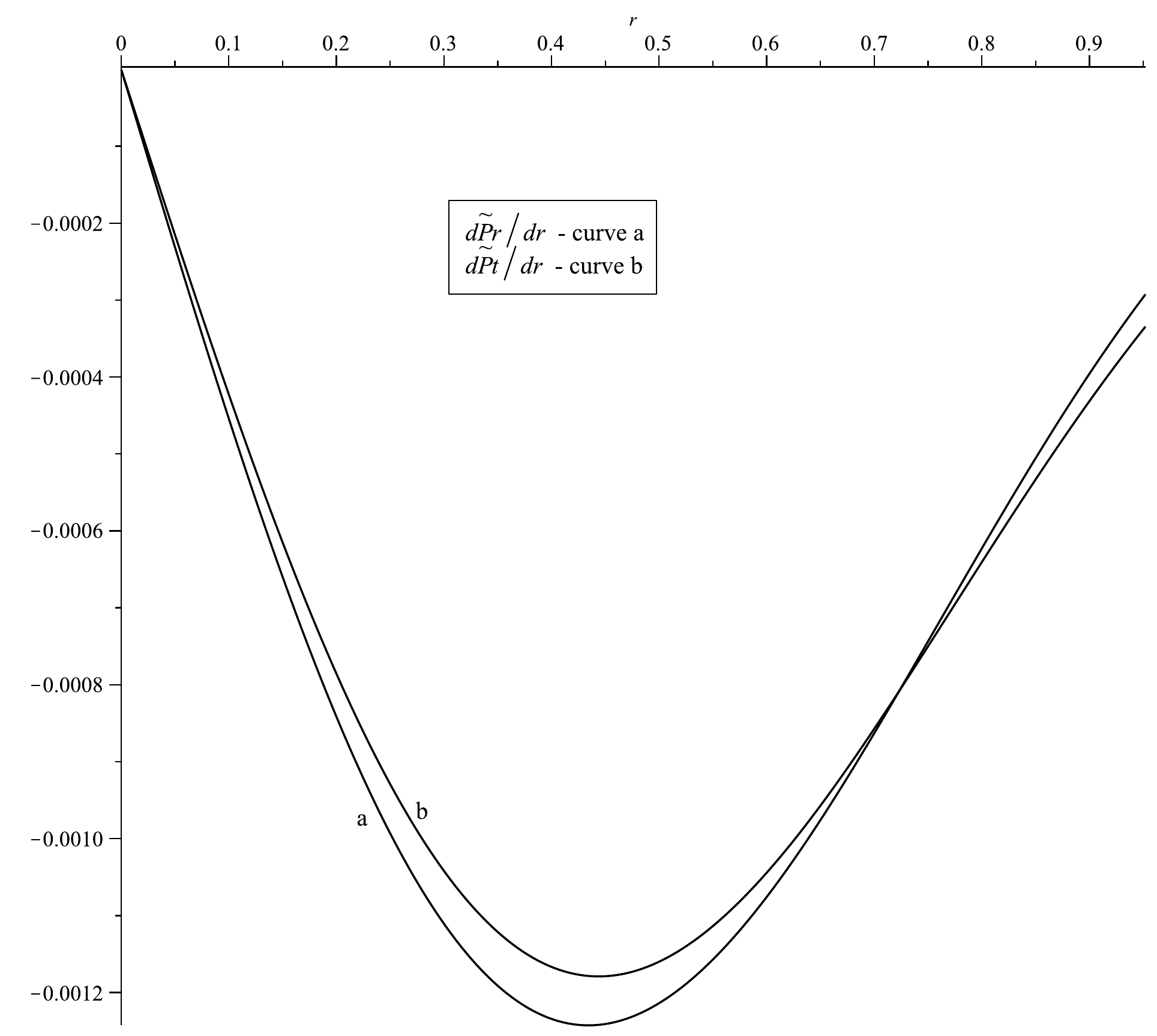}
}
\caption{Gradients pressures for the solution obtained with the second algorithm with $\tilde{F}\not=0$ form the isotropic Gold III solution.}
\label{fig:21}       % Give a unique label
\end{figure}

\begin{figure}
\resizebox{0.4\textwidth}{!}{%
  \includegraphics{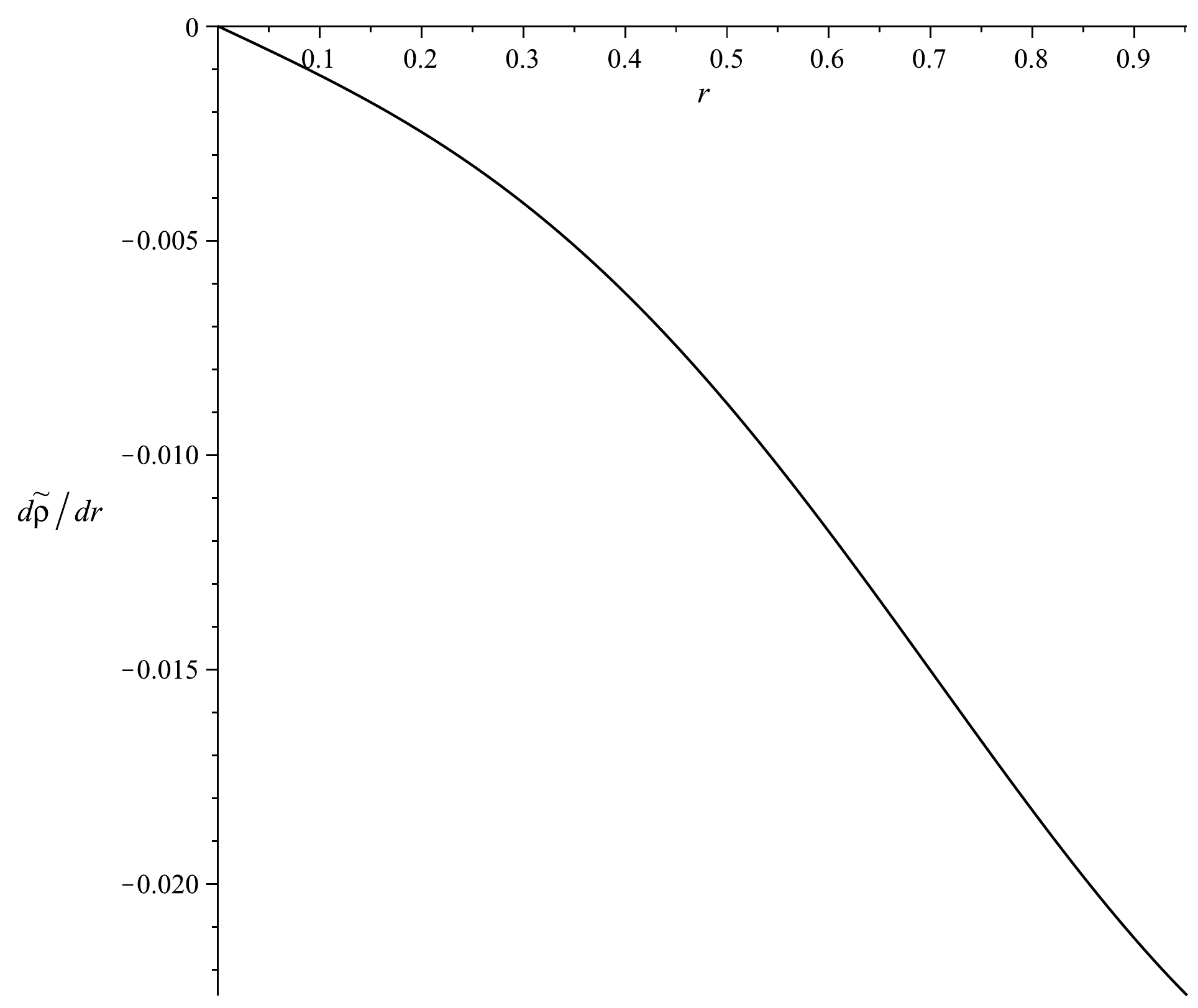}
}
\caption{Gradient of the energy density for the solution obtained with the second algorithm with $\tilde{F}\not=0$ form the isotropic Gold III solution.}
\label{fig:22}       % Give a unique label
\end{figure}

\begin{figure}
\resizebox{0.4\textwidth}{!}{%
  \includegraphics{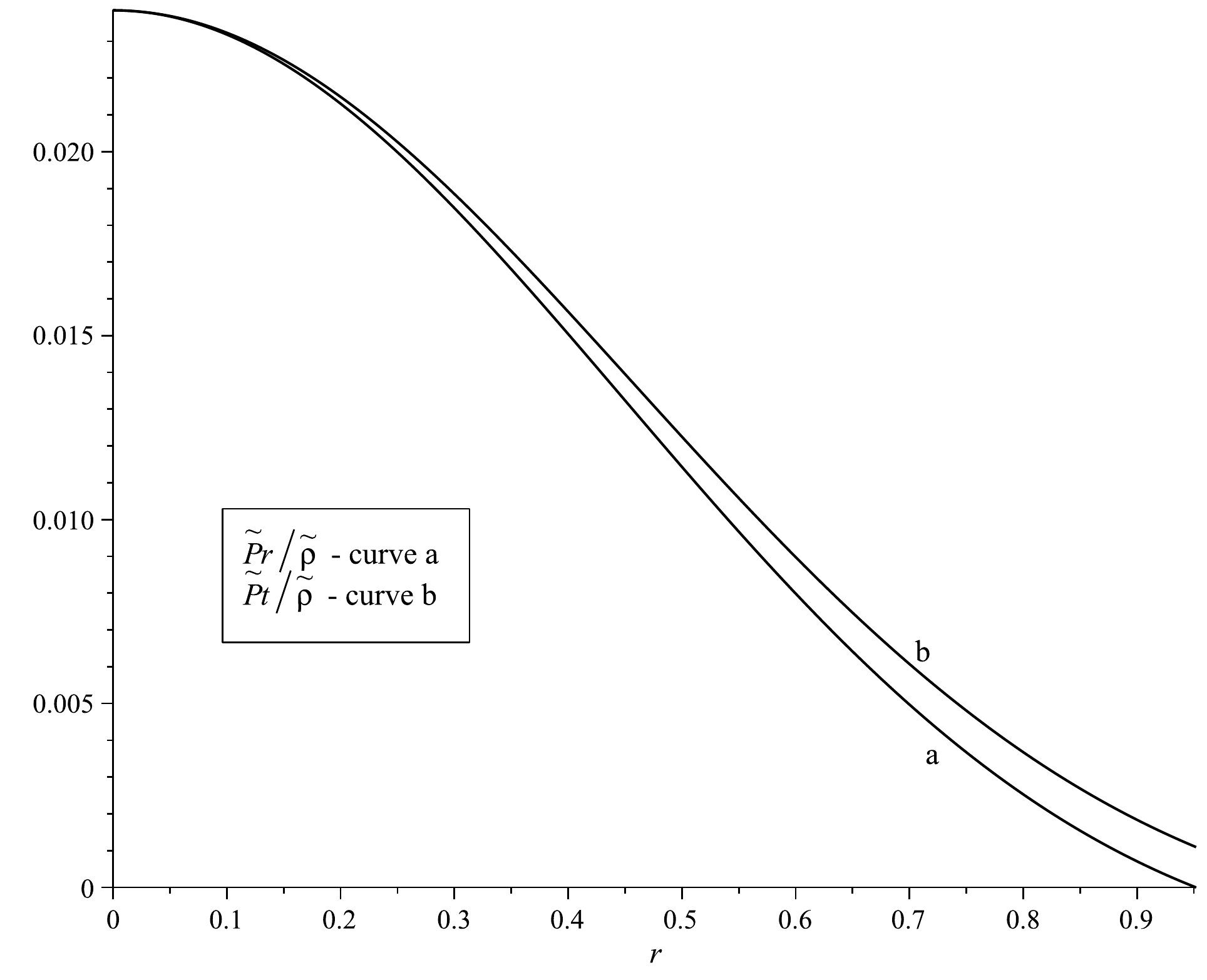}
}
\caption{Dominant energy condition for the solution obtained with the second algorithm with $\tilde{F}\not=0$ form the isotropic Gold III solution.}
\label{fig:23}       % Give a unique label
\end{figure}

\begin{figure}
\resizebox{0.4\textwidth}{!}{%
  \includegraphics{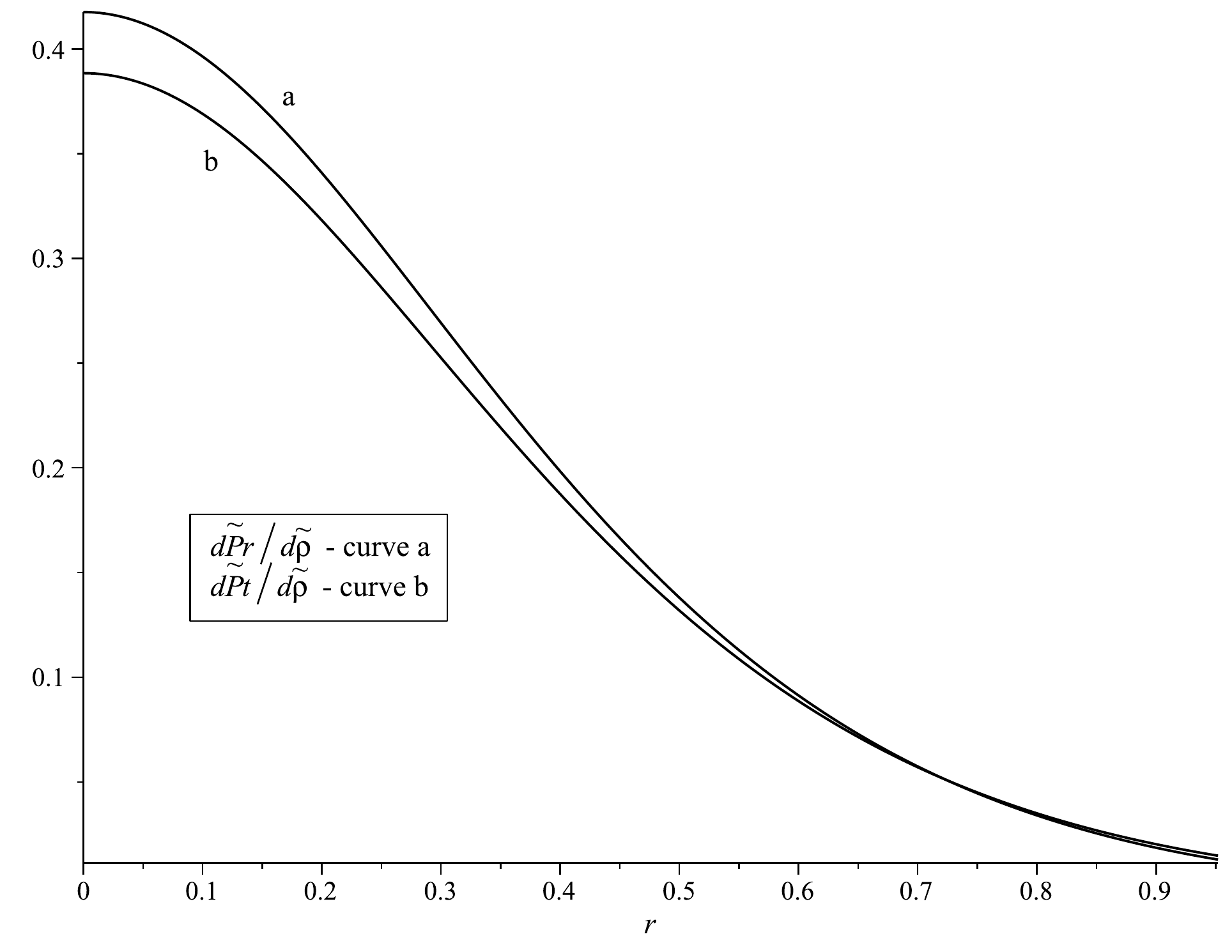}
}
\caption{Causality condition for the solution obtained with the second algorithm with $\tilde{F}\not=0$ form the isotropic Gold III solution.}
\label{fig:24}       % Give a unique label
\end{figure}

\subsubsection*{Solution 2}
In order to see if we can improve the behavior of the previous solution let us now take a more complicated case in which $\tilde{F}\not=0$. In particular we will suppose that

\begin{equation}
    \tilde{F}=\frac{4bG}{\sqrt{B}}\left(\frac{g-1}{g+1}\right)(g^2-1)^{-3/2}, \label{F}
\end{equation}
where $G$ is a constant that allow us to connect with the previous case ($\tilde{F}=0$). Then we can obtain

\begin{eqnarray}
f & = & \frac{1}{g(g+1)}\left[16CDr^6\sqrt{B}b^2 -\frac{G}{6}\right], \label{fg2} \\
J & = & -\frac{G^2[g(g+1)+2r^2b(g^2-1)^{1/2}(2g+1)]}{r^2g^2(g+1)^2}. \label{J}
\end{eqnarray}
Now, in this case the regularity in the center of the distribution implies

\begin{equation}
    B = \frac{9(e^{2a}+1)^4}{[3(e^{2a}+1)^2+2e^a(\alpha G e^a +3(e^{2a}+1))]^2}.
\end{equation}
As in the previous case, introducing (\ref{F})-(\ref{J}) in the systems (\ref{h0g})-(\ref{h2g}) and (\ref{t0g})-(\ref{t2g}) we can obtain the effective pressures and energy density from  
\begin{eqnarray}
\tilde{\rho} & = & \rho +\alpha \Theta^0_0 + \alpha^2 H^0_0, \\
\tilde{P}_r & = & P -\alpha \Theta^1_1 - \alpha^2 H^1_1 ,\\
\tilde{P}_t & = & P -\alpha \Theta^2_2 - \alpha^2 H^2_2. 
\end{eqnarray}

Finally, to give an example we choose the values $G=D$, $D\alpha=0.304$, $a=3.3$, $b=1.02$ and $C=0$. Then with the matching conditions it is easy to obtain $r_\Sigma = 0.952$, $r_{1\Sigma}= 0.913$, $M=0.047$, $D=0.953$ and $\alpha=0.319$.In this case the pressures and the energy density are in the figures (\ref{fig:19}) and (\ref{fig:20}), while the acceptability conditions are in the figures (\ref{fig:21})-(\ref{fig:24}).

\section{Conclusions}
\label{coclu}
The Minimal Geometric Deformation is an efficient method to obtain new analytical internal solutions to Einstein's equations. However, the main advantage of MGD is not present if one considers a line element in isotropic coordinates, because it is not possible to decouple the system of equations related to the gravitational sources.
 
In this work, we present two inequivalent methods to obtain analytical and anisotropical solutions of Einstein's equations in isotropic coordinates, starting with a known (isotropic or anisotropic in pressures) solution. Although this is not exactly the known Minimal Geometric Deformation, both algorithms uses the same ansatz for the metric deformations. The differences with MGD is the following: in the first method there is no decoupling of Einstein's equations and we only use convenient combinations of the equations to solve the complete system. In the second method, there is a decoupling of the system of equations. Nevertheless, to achieve this, is necessary to include an additional source ($H_{\mu \nu}$) that also have to satisfy Einstein's equations.  

In order to test the two methods we select Gold III as a seed solution and with each algorithm we extend it to the anisotropical domain using two different conditions over the new sources. Indeed, we have found four analytical solutions in isotropic coordinates, completely different from the initial ones, and with local anisotropic pressures. It is important to mention that the inequivalence of both methods is evident. This can be easily verified by selecting the same seed solution and the same values for the free parameters. This way, it can be verified that the resulting solutions are different. In fact, selecting the values for the parameters used in the second method in order to obtain a physical acceptable solution, the first method throws a solution that does not satisfy the conditions for physical acceptability.

As we can observe from figures (\ref{fig:1}) and (\ref{fig:13}), the solution obtained from Gold III with $F=0$ in the first method and $\tilde{F}=0$ in the second one exhibit regions, in the internal part of the distributions, in which the radial pressure is greater than the tangential pressure. This behavior could suggest that this models are unstable under perturbations. In fact the change of sign in the force contribution coming from the local anisotropy indicates the possible appearance of cracking (see \cite{cn,cn1,cn2,cn3,cn4,cn5,cn6,cn7}) under perturbations. It had been widely discussed in the literature that the presence of cracking can affect the evolution of compact objects in a drastically way when these depart from the equilibrium. Indeed cracking phenomenon can be associated to different scenarios that can be related with observations (for example \cite{33,34,36,37,38,39,40,41,42}). 

However, in both methods we show that this behavior can be corrected when $F\not=0$ in the first one and $\tilde{F}\not=0$ in the second one. Indeed, we show that the solutions displayed in the figures (\ref{fig:9})-(\ref{fig:12}) and (\ref{fig:19})-(\ref{fig:24}) satisfied all the physical acceptability conditions. Another possibility to avoid the odd behavior of the solutions obtained here with $F=0$ ( $\tilde{F}=0$ in the second method), which we do not discuss, consist in doing the matching with an exterior solution different from the Schwarzschild's vacuum.  

Finally, we what to mention that the complicate form that solutions have, suggest that this may be new solutions of the Einstein's equations not reported in the literature, at least compared with other solutions in isotropic coordinates. Nevertheless, to probe that these are indeed new solutions of the field equations (including the rewriting in any other coordinates) it is necessary to perform a deeper analysis (computing the scalar invariants of our solutions and comparing them with those obtained from all the others known solutions of the Einstein's equation's) which is beyond  the purpose of this work. Our main goal in this work is not to study any particular model. We only center our attention in presenting two different simple algorithms to find analytical solutions to Einstein's equations in isotropic coordinates. In order to describe any particular model, it is necessary to consider $F(r)$ (or $\widetilde{F}(r)$) related with the system under study, in order to solve the resulting equations.    

\section{Acknowledgements}
\label{agra}

We want to say thanks for the financial help received by the Projects ANT1855 and ANT1856 of the Universidad de Antofagasta. P.L wants to say thanks for the financial support received by the CONICYT PFCHA / DOCTORADO BECAS CHILE/2019 - 21190517 and also to Professor M. Asorey of the Physics Department of the Universidad de Zaragoza where part of this work was done. C.L.H wants to say thanks for the financial support received by CONICYT PFCHA / DOCTORADO BECAS CHILE/2019 - 21190263, and to Professor B. Fiol from the Physics Department of the
Universidad de Barcelona for his hospitality during the
realization of part of this work. P.L and C.L.H are also grateful with Project Fondecyt Regular 1161192 and Semillero de Investigaci\'on SEM 18-02 from Universidad de Antofagasta.

%
% BibTeX users please use
% \bibliographystyle{}
%\bibliography{biblio2}
%
% Non-BibTeX users please use

\end{document}